\documentclass[aip,jcp,amsmath,amssymb,longbibliography,reprint]{revtex4-1}

\usepackage{longtable}
\usepackage{algorithmicx}
\usepackage{algpseudocode}
\usepackage{array}
\usepackage{color}
\usepackage[T1]{fontenc}
\usepackage{float}
\usepackage{hyperref}
\usepackage{graphicx}
\usepackage[utf8]{inputenc}
\usepackage{lmodern}
\usepackage{makecell}
\usepackage[version=4]{mhchem}
\usepackage{physics}
\usepackage[normalem]{ulem}
\usepackage[dvipsnames]{xcolor}
\usepackage{xspace}
\usepackage{xfrac}

\graphicspath{{figures/}}

\newcommand{\abinitio}{\textit{ab initio}\xspace}

\newcommand{\etal}{\textit{et al.}\xspace}

\newcommand{\ang}{\text{\,\AA}\xspace}
\newcommand{\bohr}{\text{\,a$_0$}\xspace}
\newcommand{\icubang}{\text{\,\AA$^{-3}$}\xspace}

\newcommand{\wvn}{\text{\,cm$^{-1}$}\xspace}
\newcommand{\sh}{\text{-}}

\newcommand{\parahyd}{\textit{para}-{H}$_2$\xspace}

\usepackage{silence}
\begin{document}


\title{A neural network-based four-body potential energy surface for parahydrogen}

\author{Alexander Ibrahim}
\affiliation{
    Department of Physics and Astronomy,
    University of Waterloo,
    200 University Avenue West, Waterloo, Ontario N2L 3G1, Canada
}
\affiliation{
    Department of Chemistry,
    University of Waterloo,
    200 University Avenue West, Waterloo, Ontario N2L 3G1, Canada
}

\author{Pierre-Nicholas Roy}
\email{pnroy@uwaterloo.ca}
\affiliation{
    Department of Chemistry,
    University of Waterloo,
    200 University Avenue West, Waterloo, Ontario N2L 3G1, Canada
}


\begin{abstract}
We present an isotropic \abinitio (\parahyd)$_4$ four-body interaction potential energy surface (PES).
The electronic structure calculations are performed at
the correlated coupled-cluster theory level,
with single, double, and perturbative triple excitations.
They use
an atom-centred augmented correlation-consistent double zeta basis set,
supplemented by a $(3s3p2d)$ midbond function.
We use a multilayer perceptron to construct the PES.
We apply a rescaling transformation to the output energies during training
to improve the prediction of weaker energies in the sample data.
At long distances,
the interaction energies are adjusted to match
the empirically-derived four-body dispersion interaction.
The four-body interaction energy at short intermolecular separations is net repulsive.
The use of this four-body PES,
in combination with a first principles pair potential for \parahyd
[J. Chem. Phys. \textbf{119}, 12551 (2015)],
and an isotropic \abinitio three-body potential for \parahyd
[J. Chem. Phys. \textbf{156}, 044301 (2022)],
is expected to provide closer agreement with experimental results.

\end{abstract}


\maketitle


\section{Introduction}
The topic of many-body intermolecular interactions
has seen a rise in interest over the past few decades.
\cite{manybody:15medd, manybodygpu:17fan, manybody:19alka, manybody:20oter, manybody:94elro, manybody:16cisn}
Hydrogen,
as the simplest molecular species,
is an excellent subject for the study of many-body interactions.
One of its nuclear spin isomers,
parahydrogen (\parahyd)
displays a variety of exotic quantum behaviors in many-body systems.
For example,
solid \parahyd is a quantum molecular solid.
\cite{ph2solidexp:60gush, ph2solidtheo:66nosa, ph2solidexp:67bost, solidh2theo:83kran, ph2solidexp:12fern}
The individual \parahyd molecules that comprise it
have relatively large zero-point energies.
\cite{ph2solidexp:12fern, ph2solidtheo:66nosa, ph2solidtheo:17duss, ph2solidexp:80silv}
This gives each molecule very large amplitude vibrations about its nominal lattice site,
which ``inflates'' the lattice
and gives the solid a much larger molar volume than one would expect from a classical description.
\cite{ph2solidexp:80silv, pathinteg:91zopp}

It has become increasingly apparent that
condensed many-body systems of \parahyd molecules,
especially at high densities,
cannot be adequately modelled using only pairwise additive interactions.
\cite{pathinteg:95cepe, threebody:96wind, threebody:08hind, threebody:10manz}
This appears strange at first,
because three-body and higher-order many-body interactions
are strong only at very short ranges.\cite{threebody:22ibra}
However,
these many-body interactions contribute significantly at much lower densities
than one would expect classically,
mainly because each \parahyd molecule has such a delocalized probability distribution.

Introducing many-body interactions into these models is a challenging task.
Pair interactions can often be reduced to a one-dimensional function of distance.
However,
interactions between three or more particles require several parameters to properly describe,
and are necessarily multi-dimensional.\cite{manybody:94elro}
Recent developments in multiple research fronts
have made it possible to create potential energy surfaces (PES)
for many-body systems.
Advances in computational chemistry together with increased computing power
make it possible to perform
\abinitio electronic structure calculations
for large collections of particles.
Also, the widespread availability of machine learning software
makes it much easier to create higher dimensional PESs
for many-body interactions.\cite{pytorch:19pasz, tensorflow:15abad}

Consider a collection of $ N $ \parahyd molecules.
Let the coordinate $ \vb{R}_i $ describe
the position and angular orientation of the $ i^{\rm th} $ particle.
The total interaction potential energy of the system $ V_{\rm tot} $
can be expanded as a sum of many-body potentials using
\begin{multline}
    V_{\rm tot}
    =
        \sum^{N}_{i < j} V_2(\vb{R}_i, \vb{R}_j)
        +
        \sum^{N}_{i < j < k} V_3(\vb{R}_i, \vb{R}_j, \vb{R}_k)
        + \\
        \sum^{N}_{i < j < k < l} V_4(\vb{R}_i, \vb{R}_j, \vb{R}_k, \vb{R}_l)
        +
        ...
\end{multline}
\noindent
where $ V_2 $ is the two-body interaction potential,
$ V_3 $ is the three-body interaction potential,
and so on.

Most research on interaction potentials between hydrogen molecules focuses on the pair potential.
With hydrogen being the simplest molecule,
there is no shortage of examples of two-body interaction potentials.\cite{h2pes:48hirs, h2pes:54maso, h2pes:72farr, h2pes:78silv, h2pes:83buck, h2pes:84norm, h2pes:00diep, h2pes:08hind, h2pes:08patk, ph2cluster:14faru}
Pair potentials for hydrogen molecules have several common features.
They have a repulsive wall at short intermolecular distances,
a somewhat deep attractive well (typically centred around $ 3.0 \ang $),
and a weakly decaying attractive tail for the dispersion interaction.
A common issue with several \ce{H2} and \parahyd pair potentials
is that the repulsive wall is too strong.
This leads to a severe overestimation of the pressure
when modelling condensed systems of \parahyd at high densities.\cite{h2pes:13omiy, ph2solidtheo:06oper}
At short intermolecular separations,
pair potentials on their own become unreliable,
and we require many-body interactions to accurately describe collections of \parahyd molecules.
\cite{threebody:96wind, threebody:08hind, threebody:10manz}

It is well known that the next order interaction,
the three-body interaction potential for \parahyd,
is overall attractive at short intermolecular distances.\cite{threebody:96wind, threebody:08hind, threebody:10manz, threebody:22ibra}
One common three-body analytic potential is the Axilrod-Teller-Muto (ATM) potential.\cite{threebody:43axil, threebody:43muto}
It accurately models the dispersion relation at large distances,
but incorrectly predicts the short-range interaction to be net repulsive.
In 2010,
Manzhos \etal published a machine-learning based three-body PES for \parahyd.\cite{threebody:10manz}
Recently,
our group published an isotropic 3D \abinitio three-body PES for \parahyd.\cite{threebody:22ibra}
Its energies are calculated at
the correlated coupled-cluster theory level,
with single, double, and perturbative triple excitations (CCSD(T)).
The calculations use
an atom-centred augmented correlation-consistent triple zeta (AVTZ) basis set,
supplemented by a $(3s3p2d)$ midbond function.\cite{elecstr:92tao}
We use the reproducing-kernel Hilbert Space method
to fit the energies.\cite{rkhs:01holl, rkhs:03ho, rkhs:17unke, rkhs:21unke}

Recently,
we performed simulations of solid \parahyd
using the two-body Faruk-Schmidt-Hinde (FSH) potential\cite{ph2cluster:14faru}
and our recent three-body PES.\cite{pathinteg:22ibra}
Due to its repulsive character at short distances,
simulations that use only the FSH potential
overestimate the pressure-density equation of state (EOS) at high densities.\cite{pathinteg:19ibra}
Because the three-body PES is net attractive,
its inclusion alongside the pair PES
decreases the pressure.
At low densities,
this change slightly improves the agreement of the simulations with experiment.
However,
at high densities,
the three-body PES overcorrects the repulsive effects of the pair PES,
and greatly underestimates the pressure.

Wheatley \etal recently published
an \abinitio four-body PES for helium.\cite{fourbody:23whea}
The energies were calculated at the CCSD(T) level,
using an AVQZ atom-centred basis set.
The interpolation for the PES is based on Gaussian processes.\cite{gaussianprocesses:22grah}
They performed path-integral Monte Carlo (PIMC) simulations using this potential
to calculate the fourth virial coefficient of helium to reasonable agreement with experiment
over a wide range of temperatures.
The four-body PES for helium
shares many qualitative features with the four-body PES for \parahyd
presented in this paper.

In this paper,
we present a six-dimensional, isotropic four-body PES for \parahyd.
The energies are calculated at the CCSD(T) level of theory,
using an AVDZ atom-centred basis set,
with an additional $ (3s3p2d) $ midbond function
at the centre of mass of each calculation.\cite{elecstr:92tao}
We use a rigid rotor approximation
and average out the rotational degrees of freedom using the 6-point Lebedev quadrature\cite{lebedev:76lebe}
to make the interactions isotropic.
We model the PES using a multi-layer perceptron,
trained using PyTorch.\cite{pytorch:19pasz}
At large intermolecular separations,
we use an analytic four-body interaction potential
to more accurately describe the dispersion interaction.\cite{fourbody:57bade, fourbody:58bade}
At short range,
we fit the energies to an exponential decay,
with some modifications for numerical stability.

The remainder of this paper is structured as follows:
Sec.~(\ref{sec:creating_the_pes}) goes over the creation of the PES,
including the calculation of the interaction energies and
how the degrees of freedom are managed.
It also covers how the training data is generated,
the transformations on the inputs and outputs that improve
the effectiveness of the training,
and the training process itself.
Sec.~(\ref{sec:discussion_and_analysis}) is a discussion and analysis of the PES.
It includes an error analysis of the electronic structure calculations
and the methods used to manage the degrees of freedom.
It also covers the short-range and long-range extrapolations of the PES,
and ends with an analysis of the effects of the four-body interaction
on a classical approximation of solid \parahyd.

\section{Creating the Potential Energy Surface} \label{sec:creating_the_pes}
\subsection{Electronic Structure Calculations} \label{sec:energy_calculations}

We use the MRCC software program (version $2019$)\cite{elecstr:13roli}
to carry out the electronic structure calculations for the $(\ce{H2})_4$ energies.
The calculations are performed at the CCSD(T) level of theory.\cite{elecstr:89ragh}
There is an AVDZ basis set
centred at each of the eight hydrogen atoms.
We supplement the basis with $ (3s3p2d) $ midbond functions\cite{elecstr:92tao}
located at the centre of mass of the collection of atoms,
as was done for the pair potential by Hinde\cite{h2pes:08hind}
and this group's recent three-body PES.\cite{threebody:22ibra}
An analysis of different basis sets and electronic structure methods
is given in Sec.~(\ref{sec:basis_set_and_method}).

There are many settings available in the MRCC program
that determine the precision of the energy calculations.
All calculations provided in this paper were carried out
using the default settings, except for
\texttt{cctol}, which is set to 9,
and \texttt{scftol} and \texttt{scfdtol}, which are both set to 7.
Under these settings,
the final iterations in the SCF and CCSD calculations
change on the order of $ 10^{-11} $ Hartrees and $ 10^{-10} $ Hartrees, respectively.
Nearly all SCF and CCSD energy calculations converged within 9 and 20 iterations, respectively.
After the conversion to wavenumbers
and the sum of terms needed to find the interaction energy,
we estimate a conservative upper bound of $ 10^{-4} \wvn $
on the convergence error of the interaction energies.
As a test,
we select a handful of calculations for the tetrahedron geometry
(as presented in Sec.~(\ref{sec:basis_set_and_method})),
with side lengths ranging from $ 2.2 \ang $ to $ 4.5 \ang $.
We perform them once more,
but with the value of \texttt{cctol} set to 10 instead of 9.
Of the differences between the energies in our samples when \texttt{cctol=10} and \texttt{cctol=9},
the mean absolute difference was $ 7.2 \times 10^{-6} \wvn $,
whereas the maximum absolute difference was $ 2.7 \times 10^{-5} \wvn $.

Let $ \vb{R}_i = (\vb{r}_i, \rho_i, \vb{\Omega}_i) $
describe the centre-of-mass position, bond length, and space-fixed angular orientation
of the $i^{\rm th}$ hydrogen molecule, respectively.
Let $ E_{ijkl} = E_{ijkl}(\vb{R}_i, \vb{R}_j, \vb{R}_k, \vb{R}_l) $
be the CCSD(T) energy with all molecules $i$, $j$, $k$, and $l$ present.
Define $ E_{ijk} $
to be the CCSD(T) energy with only molecules $i$, $j$, and $k$ present,
but still in the full four-body basis set.
In other words,
despite the charge of molecule $l$ being missing,
the atom-centred basis of molecule $l$ is still present.
We define $ E_{ij} $ and $ E_{i} $ similarly,
in that a missing index indicates that the molecule is missing but its basis set is still present.
The counterpoise-corrected\cite{elecstr:70boys} interaction energy
of four hydrogen molecules
(with $(\vb{R}_i, \vb{R}_j, \vb{R}_k, \vb{R}_l)$ omitted for brevity)
is given by
\begin{align} \label{eq:energy_calculations:interaction_energy}
\Delta E
&= E_{1234} \nonumber \\
&- (E_{123} + E_{124} + E_{134} + E_{234}) \nonumber \\
&+ (E_{12} + E_{13} + E_{14} + E_{23} + E_{24} + E_{34}) \nonumber \\
&- (E_{1} + E_{2} + E_{3} + E_{4}) \, .
\end{align}
\noindent
Each interaction energy requires 15 separate electronic structure calculations to produce.

\subsection{Managing Degrees of Freedom} \label{sec:degrees_of_freedom}
By modelling the molecules as a collection of four rotors in free space,
we can describe the system as having
4 degrees of freedom from the bond lengths,
8 degrees of freedom from the angular orientations of the rotors,
and 6 degrees of freedom from the distances between their centres of mass,
for a total of 18 degrees of freedom.

We can use the rigid rotor approximation to fix the bond length of each molecule
to the vibrationally averaged ground state length of $ 1.449 \bohr $,\cite{threebody:08hind}
removing four degrees of freedom.

The eight degrees of freedom associated with the angular orientations
are removed by performing a 6-point Lebedev quadrature.\cite{lebedev:76lebe, lebedev:98beck, lebedev:03wang}
To perform this averaging,
we first fix in space the centres of mass of each of the four hydrogen molecules.
Next we align each rigid \ce{H2} molecule along either the space-fixed $x$-, $y$-, or $z$-axis,
giving each molecule 3 possible angular orientations.
This alignment is done independently for each of the four hydrogen molecules,
for a total of 81 different combinations of space-fixed angular orientations.
Let $ n $ label each of the 81 different combinations.
Performing the energy calculation described in Eq.~(\ref{eq:energy_calculations:interaction_energy})
for the $ n^{\rm th} $ combination gives the interaction energy $ \Delta E^{(n)} $.
The isotropic four-body interaction energy is the average of these 81 interaction energies.
\begin{equation} \label{eq:degrees_of_freedom:lebedev_average}
V_4
=
\frac{1}{81}
\sum_{n=1}^{81}
\Delta E^{(n)}( \{ \vb{R}_i \} )
\, .
\end{equation}
\noindent
where $ \{ \vb{R}_i \} = (\vb{R}_1, \vb{R}_2, \vb{R}_3, \vb{R}_4) $.
This averaging process projects out
the anisotropic components of the interaction energy.
It is an approximation to performing the full integration
over all the degrees of freedom.
There are higher levels of Lebedev quadratures
that give more accurate averages,
but require many more energy calculations to perform the average.\cite{lebedev:76lebe}
In an analysis given in Sec.~(\ref{sec:lebedev}),
we show that the next highest Lebedev quadrature give a negligible improvement in the accuracy.
\begin{figure} [ht]
        \centering
    \includegraphics[width=\columnwidth]{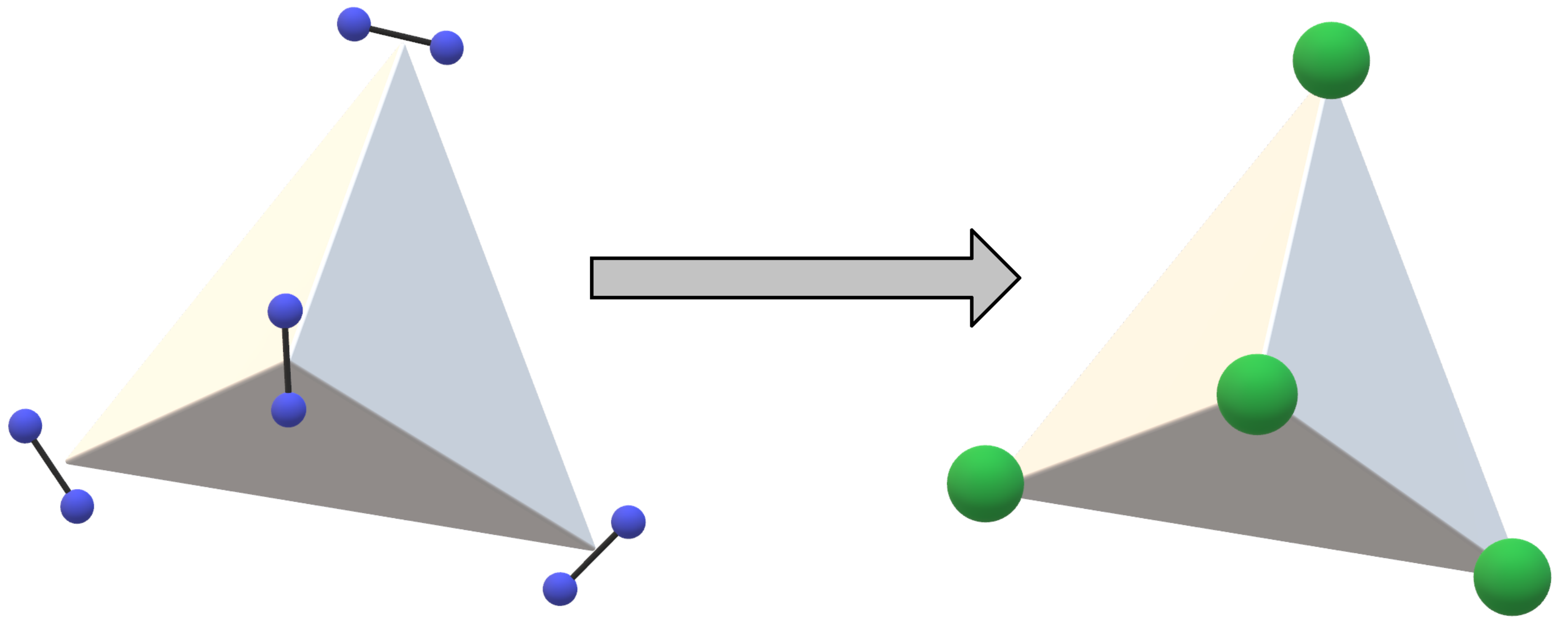}
    \caption{A graphical representation of how
        fixing the bond length of each hydrogen molecule,
        and averaging over the angular orientations of the system,
        means we now model the four hydrogen molecules as four points in space.
    }
    \label{fig:paper3_rotors_to_points}
\end{figure}

After removing the bond length and angular orientation degrees of freedom,
we have reduced the description of our system
from that of four interacting rotors
to that of four interacting points in space (see Fig.~\ref{fig:paper3_rotors_to_points}).
We are now left with only 6 degrees of freedom,
corresponding to the relative distances between the centres of mass
of each pair of hydrogen molecules.
We can thus describe each four-body geometry with the six relative side lengths
$\{ r_{ij} \} = (r_{12}, r_{13}, r_{14}, r_{23}, r_{24}, r_{34})$
where $ r_{ij} $ is the distance between the centres of mass of molecules $ i $ and $ j $.
For the remainder of the paper,
when we refer to the ``distance between two molecules,''
we mean specifically the distance between each molecule's centre of mass.
Also,
when we refer to the ``interaction energy of a four-body geometry,''
we refer to the isotropic interaction energy between four \parahyd molecules,
where the centres of mass of the molecules lie at the corners
of the aforementioned four-body geometry.

Notice that each interaction energy
is a linear combination of 15 separate CCSD(T) energies,
each requiring its own electronic structure calculation.
Moreover,
each 6-point Lebedev quadrature
is a linear combination of 81 separate interaction energies.
This means that each averaged energy value requires a total
of 1215 individual CCSD(T) energy values.

\subsection{Sample Generation} \label{sec:sample_generation}
To train the neural network and create the PES,
we need to generate samples for training, testing, and validation.
The inputs are the ordered set of six side lengths
$ (r_{12}, r_{13}, r_{14}, r_{23}, r_{24}, r_{34}) $
and the outputs are the corresponding four-body interaction energy
as given by Eq.~(\ref{eq:degrees_of_freedom:lebedev_average}).
This PES is meant to be used in numerical simulations,
including at high densities,
and thus interaction energies at short distances must be well represented in the training data.
At longer intermolecular separations,
the \abinitio CCSD(T) energies converge to the four-body dispersion relation
(see Sec.~(\ref{sec:basis_set_and_method})).\cite{fourbody:57bade, fourbody:58bade}
Thus we do not have to sample energies from geometries
with excessively large intermolecular separations,
even for purposes where the weak long-range energies are important,
such as simulations involving the gas phase.\cite{virial:13garb}

The procedure for generating samples is as follows.
First,
we decide on an exponential decay sampling function
\begin{equation} \label{eq:exponential_decay_sampling_function}
    p(r) = Q \exp \{ - Cr \} ,
\end{equation}
\noindent
where $ C $ is a positive constant
and $ Q $ is a constant chosen to normalize the distribution
between $ r_{\rm min} $ and $ r_{\rm max} $ (mentioned later).
To create an input data sample,
we first sample 6 side lengths $i.i.d.$ from $ p(r) $ between $ r_{\rm min} $ and $ r_{\rm max} $,
and label them $ (r_{12}, r_{13}, r_{14}, r_{23}, r_{24}, r_{34}) $.
We next attempt to construct a four-body geometry from the six side lengths
via trilateration.\cite{trilateration:86fang}
However, not every ordered set of 6 side lengths can be turned into a valid 3D shape.
If the 6 side lengths form a valid four-body geometry, we accept the sample.
If not,
we generate an entirely new ordered set of 6 side lengths,
and repeat the process of attempting to construct a valid 3D shape from the side lengths
until it succeeds.
In the supplementary material,
we provide an example of the function we use
to generate the four points from the six side lengths.

There are several reasons for settling upon this sampling strategy.
First, it allows us to generate samples with different, uncorrelated side lengths,
which is important because we do not know \textit{a priori}
the probability distribution of the \parahyd molecules
in the situation the PES is used in.
Second,
an exponential decay function allows us to generate
samples to have shorter side lengths with a greater frequency than longer side lengths.
This is desirable,
because samples with excessively long pair distances
have energies that are either negligibly weak
or can be replaced with the analytic four-body dispersion interaction
described in Sec.~(\ref{sec:bade_potential}).
By choosing the value of $ C $,
we can determine the distribution of side lengths,
and therefore roughly determine the energies.
For example,
a large value of $ C $ causes Eq.~(\ref{eq:exponential_decay_sampling_function})
to generate smaller side lengths,
which correspond to samples with larger energies.
Looking at Fig.~\ref{fig:rescaling:paper3_dynamic_range_of_outputs},
it appears that we were successful in generating samples with
a large range of energies,
with a slight bias towards high-energy samples.

With $ C_0 = 0.9209 \ang^{-1} $,
we sample
$ 9000 $ points between $ 2.2 \ang $ and $ 4.5 \ang $ using $ C = C_0 $
(for a wide range of side lengths and energies),
$ 1500 $ points between $ 2.2 \ang $ and $ 4.5 \ang $ using $ C = 3 C_0 $
(for samples with short side lengths and large energies),
$ 1500 $ points between $ 2.2 \ang $ and $ 4.5 \ang $ using $ C = 6 C_0 $
(for samples with very short side lengths and very large energies),
and $ 4000 $ points between $ 2.8 \ang $ and $ 4.5 \ang $ using $ C = C_0 / 6 $
(for samples with long side lengths and weak energies).

We supplement the training data using samples
that are based specifically on the geometries of the \textit{hcp} lattice.
To generate these samples,
we first select a reference molecule in the lattice.
Then we consider all four-body geometries involving this reference molecule
that satisfy two conditions;
they have at least one side length equal to the lattice constant,
and they have no side length that is more than twice the lattice constant.
Geometries that disobey these conditions have a nearly negligible contribution
to the total interaction energy,
even at a lattice constant as low as $ 2.2 \ang $.
There are $ 83 $ different four-body shapes that satisfy these conditions.
For each shape,
we calculate the interaction energy for $ 47 $ different lattice constants,
evenly spaced between $ 2.2 \ang $ and $ 4.5 \ang $,
for a total of $ 3901 $ additional samples.
A rescaling strategy for training described in Sec.~(\ref{sec:rescaling})
requires us to discard many of these \textit{hcp} lattice-based samples,
resulting in only $ 1610 $ of them being included in the training data.

The aforementioned sampling strategy for geometries has certain limitations.
The range of side lengths is relatively constrained to limit the sampling of
geometries with especially weak energies.
This means that,
within any given sample,
the largest side length is at most about twice the length
of any other side length.
This leads to relatively ``compact'' geometries.
Certain types of geometries,
such as planar or linear geometries,
are poorly represented,
and if they appear at all,
it is due to their presence in the \textit{hcp} lattice-based samples.

\subsection{Feature Transformations} \label{sec:feature_transform}
For our neural network,
the input data is the ordered set of six side lengths,
\begin{equation} \label{eq:six_side_lengths}
    (r_{12}, r_{13}, r_{14}, r_{23}, r_{24}, r_{34}) \, ,
\end{equation}
and the output is the corresponding interaction energy,
\begin{equation} \label{eq:six_side_lengths_energy}
    V_4 = V_4(r_{12}, r_{13}, r_{14}, r_{23}, r_{24}, r_{34}) \, .
\end{equation}

To simplify training and improve the performance of our model,
we apply three transformations to the six side lengths.
First, we take the reciprocals of each of the six side lengths.
Shorter pair distances generally correspond to greater energies,
and taking the reciprocal of the distance
turns smaller distances into larger input features.
Next we linearly map all the features onto $ (0, 1] $.
We then perform a permutation of the six features
in a way that preserves the symmetry of the four-body system.

Because the four \parahyd molecules are identical,
our model must be invariant to the permutation of the molecule indices.
However,
swapping indices changes the order of the six side lengths.
If, for example, we swap particle indices 1 and 2
for the six side lengths in Eq.~(\ref{eq:six_side_lengths}),
the input becomes
$(r_{12}, r_{23}, r_{24}, r_{13}, r_{14}, r_{34})$.
The neural network thus interprets the permuted input as a completely different sample,
and produces a different output energy.
On a neural network trained on data without accounting for this permutation symmetry,
an index swap can even change the output energy by tens of wavenumbers at short ranges.

Not all permutations are possible.
Of the $ 6! = 720 $ permutations of these elements, only 24 of them are allowed.
For example,
there is no way to swap the indices
such that we only interchange the first two side lengths $ r_{12} $ and $ r_{13} $,
while leaving the other four side lengths unchanged.
In particular,
we cannot simply sort the six side lengths as a way of solving this permutation issue.
As seen later in Table~\ref{tab:solid_parahydrogen:geometry_contributions},
it is possible to have two separate four-body geometries
with the same six pair distances in different, permutationally-incompatible orders.
In such a case,
naively sorting the pair distances would make the two samples indistinguishable.
In addition,
not every six-tuple input of pair distances corresponds to a valid four-body geometry,
and a naive sort could generate such six-tuples.

We decide to account for the permutation symmetry at the input transformation level
by selecting the ``smallest'' permutation,
as defined by lexicographic ordering (see the supplementary material).
For readers familiar with the Python programming language,
this is the kind of ordering that Python uses
to compare two tuples of equal size of floating-point numbers.
We have a slower, naive implementation,
and a much faster lookup-based implementation.

In the slower naive implementation,
we loop over all
24 allowed permutations of the input (after the first two transformations),
and select the one with the lowest lexicographic order.
However,
under certain conditions,
we can come up with a much faster implementation.
In the cases where the shortest and second-shortest pair distances are unique,
the permutation with the lowest lexicographic order can be uniquely determined using a lookup table.
In particular,
in a molecular simulation
the six pair distances are floating point numbers
calculated from four points in space that have been perturbed from their original positions.
Thus, in this case, the six pair distances are almost always unique,
and this much faster solution becomes viable.

Another considered solution was to use the permutations for data augmentation.
However,
as seen in Sec.~(\ref{sec:neural_network}),
we propose the possibility of using small models during the simulation
for improved runtime performance.
When we make the model permutation invariant,
the increased error from using a smaller model is easily manageable.
However,
when we train using permutations for data augmentation,
the error from smaller models increases considerably,
because there are too few parameters to fit the permutations.

\subsection{Rescaling the Energies} \label{sec:rescaling}
The energies in the overall data set span about five orders of magnitude.
Our model must be able to predict both
short-range, strong-energy samples and long-range, weak-energy samples
with high accuracy.
Despite being much smaller in magnitude,
at lower densities,
the probability distribution of solid parahydrogen is predominantly
in the long-range, weak-energy region of coordinate space.
The mean-squared error (MSE) loss function does not prioritize the prediction
of weak-energy samples during training.
Other loss functions that might work well with outputs over a large range of values,
such as the relative squared error, or the mean square log error,
were considered,
but provided unsatisfactory results.

Instead,
we can take advantage of the analytically predicted behavior of the interaction potential
to transform the energies in a way that makes training more effective.
We create a rescaling function for the energies.
This rescaling function is the sum of an exponential decay and an $ r^{-12} $ decay function
\begin{equation} \label{eq:rescaling:rescaling_function}
    \phi(r) = A \exp{- B r} + C r^{-12}
\end{equation}
\noindent
where
$ A = 3.1803 \times 10^{6} \wvn $,
$ B = 4.623057 \ang^{-1} $,
$ C = 4220.011 \ang^{12} \wvn $, and
$ r $ is the mean of the six side lengths of a given sample.
The choice of using an exponential decay and a $ r^{-12} $ power decay
was to make the rescaling function somewhat resemble the features
of the true underlying \abinitio potential (see Sec.~(\ref{sec:bade_potential})).
This makes it easier to rescale the energies into a similar range.
Importantly,
the rescaling function is positive for
the entire coordinate space of the samples.
This means we can divide the \abinitio energies by this rescaling function
to recover a rescaled energy.
The neural network is trained using these rescaled energy values as outputs.
To recover the predicted energy from the output of a neural network,
we multiply the output of the neural network by the rescaling function Eq.~(\ref{eq:rescaling:rescaling_function}).

Certain weak-energy samples with large intermolecular separations
from the \textit{hcp} data set
end up becoming very large when rescaled,
due to the $ r^{-12} $ term in the denominator.
These samples have large enough side lengths
that their energies will be completely overwritten by the Bade potential
in the final PES anyways (see Sec.~(\ref{sec:basis_set_and_method})).
By filtering out all samples where the mean side length is greater than $ 4.5 \ang $,
we can remove the outsized influence of these samples on the loss function,
and improve our rescaling strategy.
This decreases the number of training samples from the \textit{hcp} data set from $ 3901 $ to $ 1610 $.

\begin{figure} [ht]
        \centering
    \includegraphics[width=\linewidth]{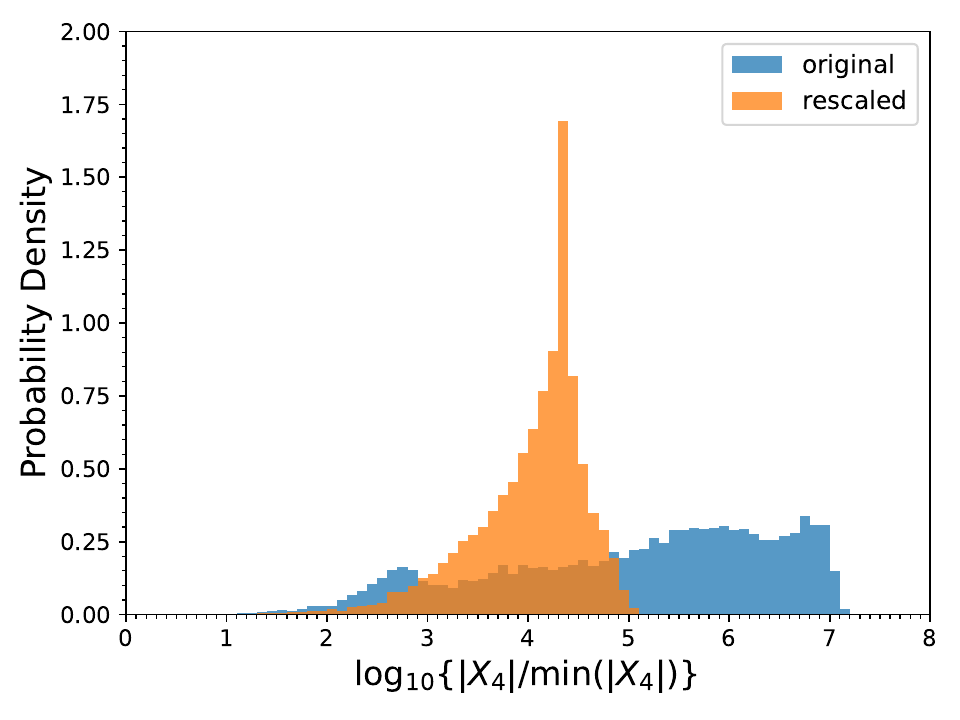}
    \caption{
        Normalized histograms of the sample outputs,
        after dividing them by the smallest value,
        then taking the absolute value followed by the logarithm.
        The variable $ X_4 $ is a stand-in for either
        the four-body interaction potential energy $ V_4 $ (blue) or
        the rescaled energy $ V_4 / \phi $ (orange),
        where $ \phi $ is given by Eq.~(\ref{eq:rescaling:rescaling_function})
        and is evaluated at the appropriate coordinates.
    }
    \label{fig:rescaling:paper3_dynamic_range_of_outputs}
\end{figure}

The rescaling operation decreases the dynamic range of the outputs.
To show this,
we first take the output energies (after the aforementioned filtering),
and divide them all by the sample with the smallest absolute energy.
We then take the logarithm of the absolute value of the results.
The same operations are performed on the rescaled energies.
Normalized histograms of the results are shown
in Fig.~\ref{fig:rescaling:paper3_dynamic_range_of_outputs}.
Of the original energies,
about $ 95 \% $ of the samples lay within a range of five orders of magnitude.
After the rescaling,
about $ 94 \% $ of the samples lay within a range of only two orders of magnitude.
Reducing the dynamic range of the output values
greatly improves the model's prediction of the weak energy samples,
which are important at lower densities.

\subsection{Neural Network and Training} \label{sec:neural_network}

We train five neural networks in total.
Each is a simple multilayer perceptron,
created, trained, and evaluated using PyTorch.\cite{pytorch:19pasz}
One of the intended uses for this PES is for Monte Carlo simulations
of large collections of \parahyd molecules,
where the evaluation of the four-body potential likely dominates the simulation's runtime.
In the recently published paper by Wheatley \etal on the four-body PES for helium,\cite{fourbody:23whea}
they noted that ther four-body potential requires much more computing time than
the two-body or three-body potentials,
which limited the temperatures for which they could perform their PIMC simulations.
The model's runtime performance is thus an important factor,
and we make reasonable trade-offs concerning other aspects of the network to accommodate it.
For example,
in addition to large networks,
we also create smaller networks that are faster to evaluate,
even though they may have larger testing errors.

\begin{figure} [ht]
        \centering
        \includegraphics[width=\columnwidth]{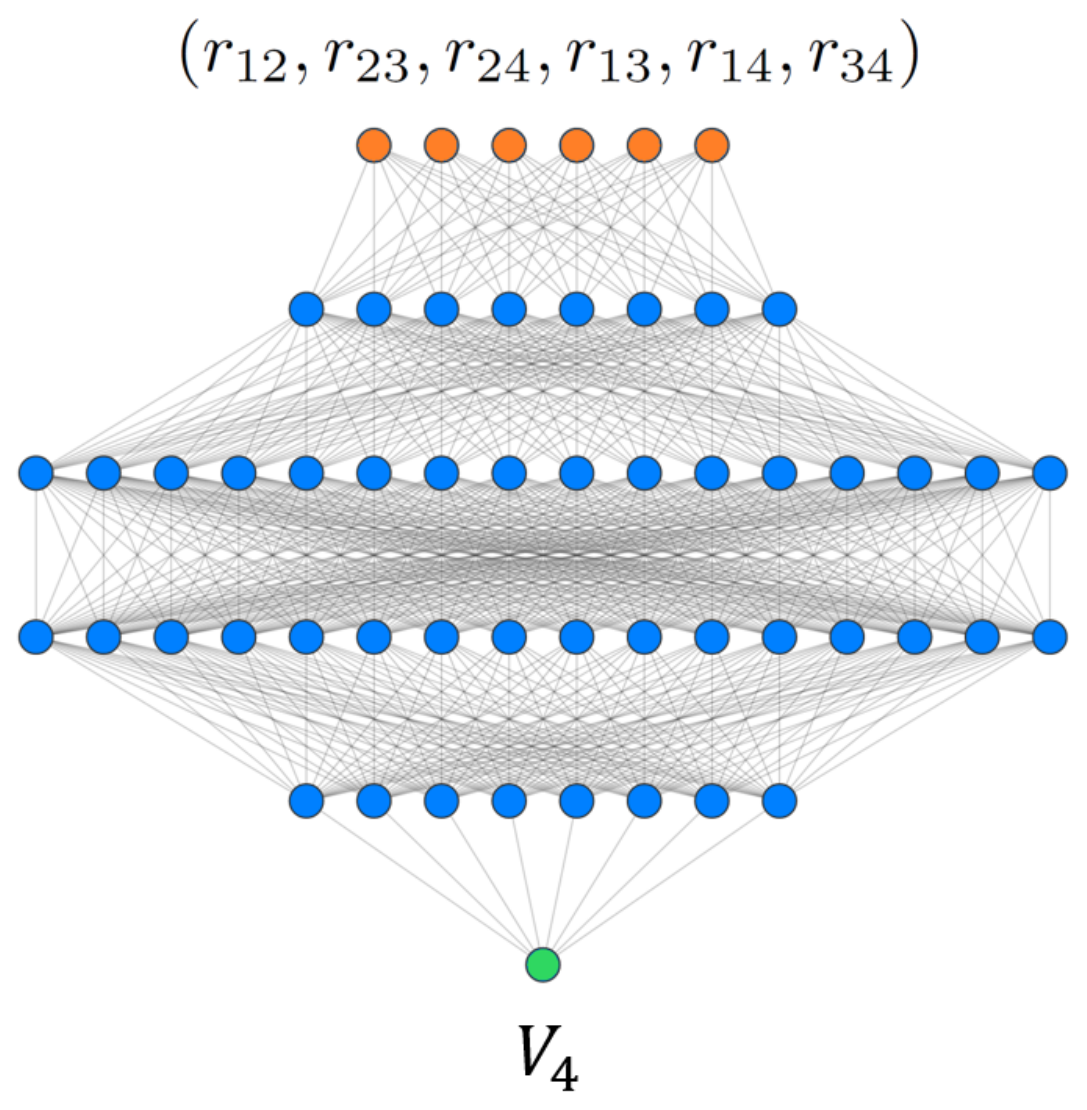}
        \caption{
    A depiction of the $ 8\sh16\sh16\sh8 $ model.
    The six orange circles represent the input features.
    The blue circles represent the fully-connected linear hidden layers.
    The green circle represents the output.
    There are a total of 4 ReLU layers, one after each linear layer.
    The results of the calculations propagate downwards in the figure.
    Not shown are the feature transformations performed on the six side lengths,
    or the rescaling transformation performed on the output.
    Diagram made with the assistance of Ref.~[\onlinecite{nndiagram:19lena}].
    }
    \label{fig:neural_network:paper3_feedforward}
\end{figure}

We create four multilayer perceptrons of different sizes.
Each has four hidden layers,
alongside the input layer of size $ 6 $ and the output layer of size $ 1 $.
We place ReLU activation functions\cite{relu:75fuku} between the hidden layers.
The four networks are the $ 64\sh128\sh128\sh64 $ model
(where the first hidden layer has a size of $ 64 $,
the second has a size of $ 128 $, and so on),
the $ 32\sh64\sh64\sh32 $ model,
the $ 16\sh32\sh32\sh16 $ model,
and the $ 8\sh16\sh16\sh8 $ model.
A depiction of the $ 8\sh16\sh16\sh8 $ model
is given in Fig.~\ref{fig:neural_network:paper3_feedforward}.

We also train a version of the largest network
using shifted softplus\cite{schnet:18schu} activation functions between the layers,
which we call the $ 64\sh128\sh128\sh64\sh\text{SSP} $ model.
Unlike the networks trained with ReLU activation functions,
this model is continuously differentiable,
for situations where forces are required.

The $ 16000 $ \abinitio samples from the probability distributions
are split randomly into training, validation, and testing sets
using a $ 6:1:1 $ split,
giving $ 12000 $ training samples, $ 2000 $ validation samples, and $ 2000 $ testing samples.
The training data is further augmented
with the remaining $ 1610 $ \abinitio samples from the \textit{hcp} lattice geometries,
\textit{i.e.} those remaining after filtering out all samples with a mean side length greater than $ 4.5 \ang $.
The input features of each sample are passed through the transformations described in
Sec.~(\ref{sec:feature_transform}),
and the output energies are rescaled using the transformation described in
Sec.~(\ref{sec:rescaling}).

For training,
we use a batch size of 64,
a mean squared error loss function,
and an Adam optimizer with a learning rate of $ 2 \times 10^{-4} $,
supplemented with a learning rate scheduler that
decreases the learning rate by $ 1 \% $ every $ 25 $ epochs past the first $ 100 $ epochs.
The $ 64\sh128\sh128\sh64 $ and $ 64\sh128\sh128\sh64\sh\text{SSP} $ models were trained for $ 20000 $ epochs,
and the other three were trained for $ 10000 $ epochs.
We find that omitting weight decay entirely improves the model's accuracy,
most likely due to the low-noise nature of the electronic structure energies.
Both batch normalization and dropout were found to have a negligible effect on training,
and so were omitted for the final model.

\begin{figure} [ht]
        \centering
    \includegraphics[width=\linewidth]{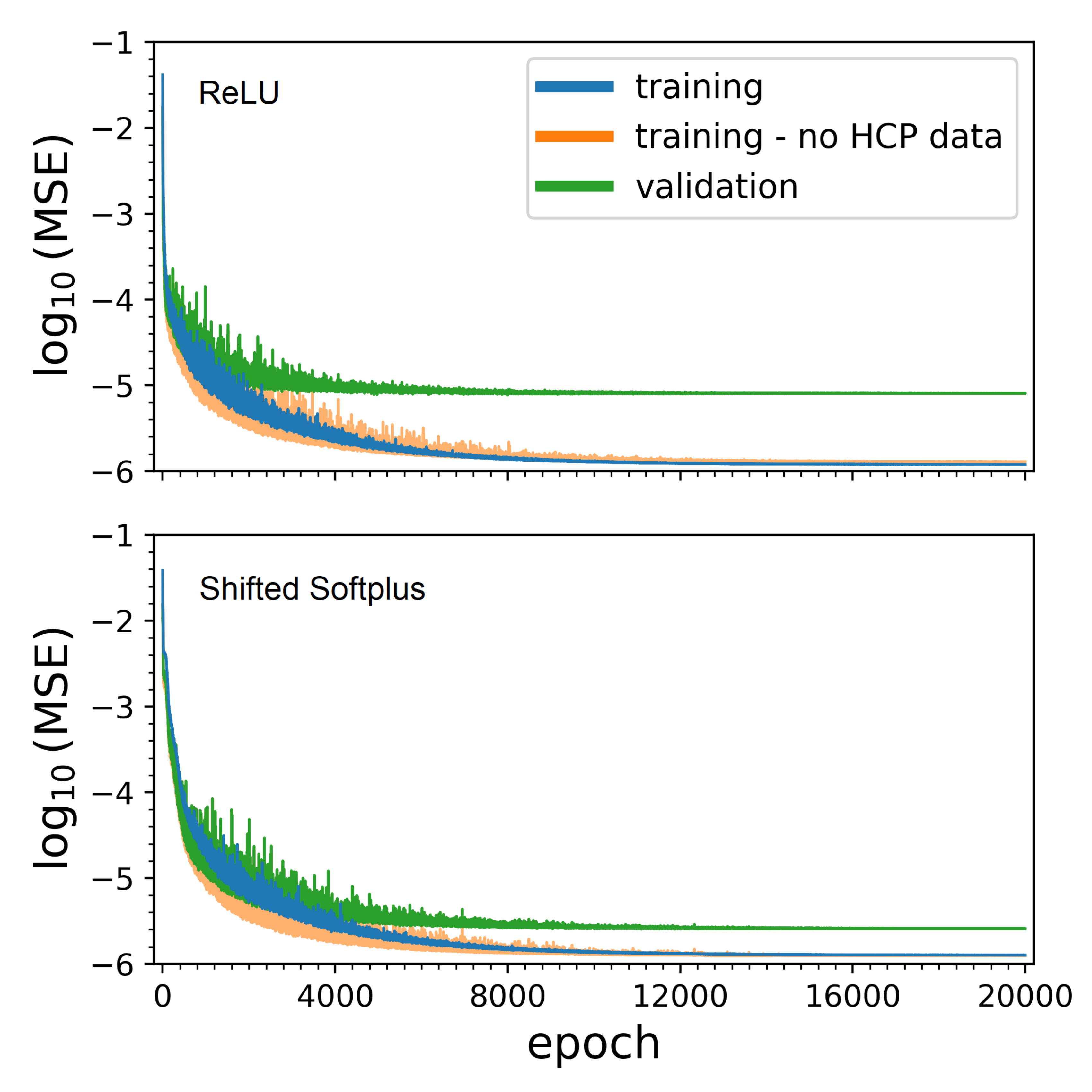}
    \caption{
    The base-10 logarithm of the mean squared error loss
    as a function of the epoch
    during the training of the $ 64\sh128\sh128\sh64\sh $ model (top),
    and the $ 64\sh128\sh128\sh64\sh\text{SSP} $ model (bottom).
    The former uses the ReLU activation function,
    while the latter uses a shifted softplus activation function.
    The plots are shown for the
    training data set (blue, solid),
    the validation data set (green, solid),
    and the training data set with the 1610 samples from
    the \textit{hcp} lattice removed (orange, semi-translucent).}
    \label{fig:neural_network:paper3_neural_network_error_64_128_128_64_both_far}
\end{figure}

In Fig.~\ref{fig:neural_network:paper3_neural_network_error_64_128_128_64_both_far},
we show the loss as a function of the epoch for the training of the $ 64\sh128\sh128\sh64 $
and $ 64\sh128\sh128\sh64\sh\text{SSP} $ models.
Shown are the results for
the training data set,
the validation data set,
and the training data set with the $ 1610 $ samples from the \textit{hcp} lattice removed.
In the supplementary material,
we show similar figures for the other, smaller models.

For both models shown in Fig.~\ref{fig:neural_network:paper3_neural_network_error_64_128_128_64_both_far},
the MSE loss decreases rapidly for all data sets.
The large amount of noise in the curves for the earlier epochs
can be attributed to the fact that we are viewing the logarithm of the MSE loss.
As the scheduler decreases the learning rate,
all error curves gradually settle.
Both models reach a similar training error,
but the use of the shifted softplus activation function
gives a lower validation error.

The models have a more difficult time fitting the \textit{hcp} lattice samples.
In both the $ 64\sh128\sh128\sh64 $ and $ 64\sh128\sh128\sh64\sh\text{SSP} $ models,
the error of the training data
exceeds the error of the training data with the \textit{hcp} lattice samples removed,
at earlier epochs.
The smaller models have even more trouble fitting the \textit{hcp} samples.
In the $ 16\sh32\sh32\sh16 $ model,
the training and validation errors are nearly the same,
and in the $ 8\sh16\sh16\sh8 $ mode,
the training error is \textit{higher} than both
the validation error and the error of the training samples with the \textit{hcp}-based samples removed.

At late stage training,
the training error becomes nearly constant,
and essentially all models beyond a certain epoch can be considered ``ideal.''
To avoid biasing the models to the test data,
we arbitrarily choose each model's weights corresponding to the very last epoch.
The root mean squared errors (RMSE) of the models when compared against the test samples,
starting with the $ 64\sh128\sh128\sh64 $ and decreasing in size,
are
$ 0.120 \wvn $,
$ 0.254 \wvn $,
$ 0.705 \wvn $, and
$ 1.495 \wvn $.
The RMSE of the $ 64\sh128\sh128\sh64\sh\text{SSP} $ model is $ 0.066 \wvn $.
As expected,
the smaller models produce worse predictions.
Using shifted softplus over ReLU as an activation function
for the largest model size
roughly halves the RMSE.

    \begin{figure} [ht]
        \centering
    \includegraphics[width=\linewidth]{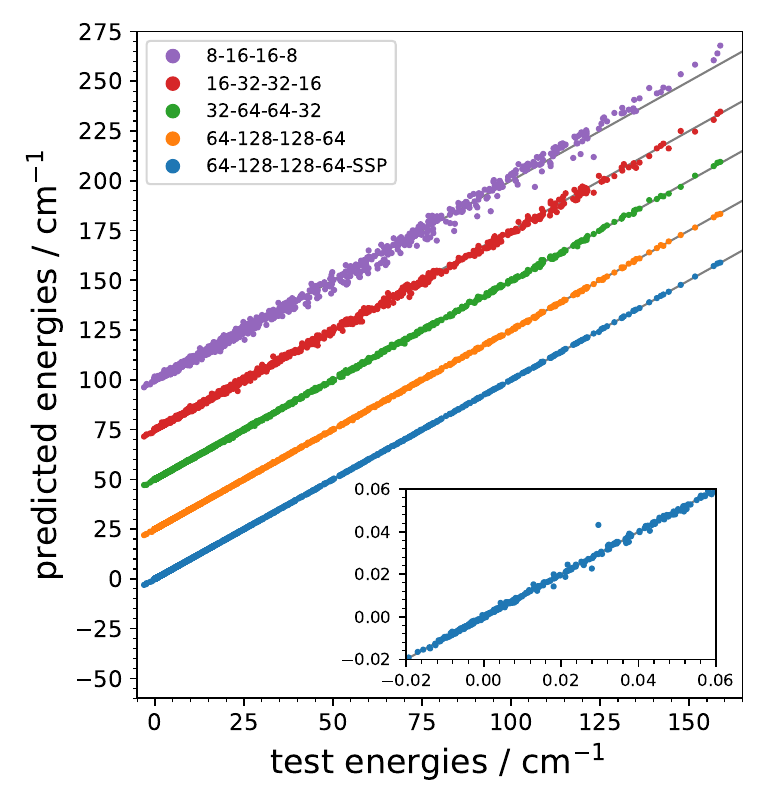}
    \caption{The four-body interaction energy predicted by each model,
    plotted against the corresponding true energy,
    for all samples in the test data set.
    To avoid overlapping the results,
    constant offsets are added to the predicted energies of each model.
    Shown are the outputs for
    the $ 64\sh128\sh128\sh64\sh\text{SSP} $ model (blue) [no offset],
    the $ 64\sh128\sh128\sh64 $ model (orange) [offset by $ 25 \wvn $],
    the $ 32\sh64\sh64\sh32 $ model (green) [offset by $ 50 \wvn $],
    the $ 16\sh32\sh32\sh16 $ model (red) [offset by $ 75 \wvn $], and
    the $ 8\sh16\sh16\sh8 $ model (purple) [offset by $ 100 \wvn $].
    The inset shows the predictions for the $ 64\sh128\sh128\sh64\sh\text{SSP} $ model
    for a certain range of weaker energies (both axes in units of $ \wvn $).
    }
    \label{fig:neural_network:model_prediction_line_all_models}
\end{figure}

In Fig.~\ref{fig:neural_network:model_prediction_line_all_models},
we show the four-body interaction energy predicted by each model,
against the corresponding true energy,
for all samples in the test data set.
The closer a point is to the diagonal line,
the better the prediction.
We see that the relatively large $ 64\sh128\sh128\sh64 $ and $ 64\sh128\sh128\sh64\sh\text{SSP} $ models
do an excellent job of predicting the interaction energies in the test set.
As we move to smaller models,
the predictions become worse.
In the inset of Fig.~\ref{fig:neural_network:model_prediction_line_all_models},
we provide a closer look at the predictions of the $ 64\sh128\sh128\sh64\sh\text{SSP} $ model
between $ -0.02 \wvn $ and $ 0.06 \wvn $.
Notably,
these predictions are of energies below the model's RMSE of $ 0.066 \wvn $.
Training the models without the energy rescaling procedure
described in Sec.~(\ref{sec:rescaling})
results in much worse predictions of the weak energy samples.

\section{Discussion and Analysis} \label{sec:discussion_and_analysis}
\subsection{Lebedev Quadrature Comparison} \label{sec:lebedev}
To create the isotropic interaction energies in our PES,
we use a 6-point Lebedev quadrature to average over
the rotational degrees of freedom of each \parahyd molecule.
\cite{lebedev:76lebe, lebedev:98beck, lebedev:03wang}
In a Lebedev quadrature,
we calculate the interaction energies of the hydrogen molecules
in specific combinations of angular orientations,
then perform a weighted average of these energies.
This weighted average only approximates
the true integration over all the rotational degrees of freedom.
The 6-point quadrature is described in Sec.~(\ref{sec:degrees_of_freedom}).

Using a larger Lebedev quadrature scheme
increases the accuracy of the approximation,
but also increases the number of terms in the sum.
The 6-point Lebedev quadrature scheme is the smallest possible scheme.
It uses only 3 angular orientations per molecule,
and with 4 molecules,
we need $ 3^4 = 81 $ energies to perform a single average.
The next largest Lebedev scheme is the 14-point scheme.
It uses 7 angular orientations per molecule,
and thus requires $ 7^4 = 2401 $ energies to perform a single average,
nearly a 30-fold increase over the 6-point scheme.

To estimate the integration error,
we select three tetrahedra of different side lengths,
and calculate the four-body interaction energies for each one,
using both the 6-point and 14-point Lebedev quadrature schemes.
The results are shown in Table~\ref{tab:lebedev:averaged_energies}.
In all three cases,
the difference between the 6-point and 14-point averaged energies
is acceptably small.
Thus, the 6-point Lebedev quadrature scheme should be acceptable
for the isotropic energies of our PES.

\begin{table} [H]
    \centering
    \caption{
        Interaction energies of four \parahyd molecules in a tetrahedron geometry
        for three chosen side lengths $ r $,
        spherically averaged using
        the 6-point ($ V_4^{\rm 6p} $)
        and 14-point ($ V_4^{\rm 14p} $)
        Lebedev quadratures.
        Calculations are performed using an AVDZ basis set at the CCSD(T) level.\label{tab:lebedev:averaged_energies}
    }
    \begin{tabular}{lcccc}
        \hline
        \hline
    
            \mbox{$ r (\ang) $} & \mbox{$ V_4^{\rm 6p} (\wvn) $} & \mbox{$ V_4^{\rm 14p} (\wvn) $} & \mbox{difference $(\%)$} \\
            \hline
            $ 2.20            $ & $  1.9917 \times 10^{2}      $ & $  1.9887 \times 10^{2}       $ & $ 0.15            $ \\
            $ 2.95            $ & $  2.3478 \times 10^{0}      $ & $  2.3484 \times 10^{0}       $ & $ 0.028           $ \\
            $ 4.00            $ & $ -5.9113 \times 10^{-3}     $ & $ -5.8332 \times 10^{-3}      $ & $ 1.3             $ \\
            \hline
            \hline
    \end{tabular}
\end{table}

\subsection{Basis Set and Method Comparison} \label{sec:basis_set_and_method}
The \abinitio energies in the PES
are calculated at the CCSD(T) level of theory,
using an AVDZ atom-centred basis
with a supplementary $(3s3p2d)$ midbond basis set at the system's centre of mass.
Our choice
of the calculation method and the size of the basis set
takes into account both
the quality of the resulting energies
and the feasibility and expense of performing the calculations.

To compare the quality of calculations for different methods and basis sets,
we calculate the four-body interaction energy of the tetrahedron geometry
for side lengths between $ 2.2 \ang $ and $ 4.5 \ang $.
We do so for four different cases:
at the HF level of theory using an AVDZ basis,
at the MP2 level of theory using an AVDZ basis,
at the CCSD(T) level of theory using an AVDZ basis,
and at the CCSD(T) level of theory using an AVTZ basis.
All calculations also use a supplementary $(3s3p2d)$ basis set at the system's centre of mass.
We also show the four-body component
of the four-body dispersion interaction potential,
also known as the Bade potential (see Sec.~(\ref{sec:bade_potential})).

\begin{figure} [ht]
    \centering
    \includegraphics[width=\linewidth]{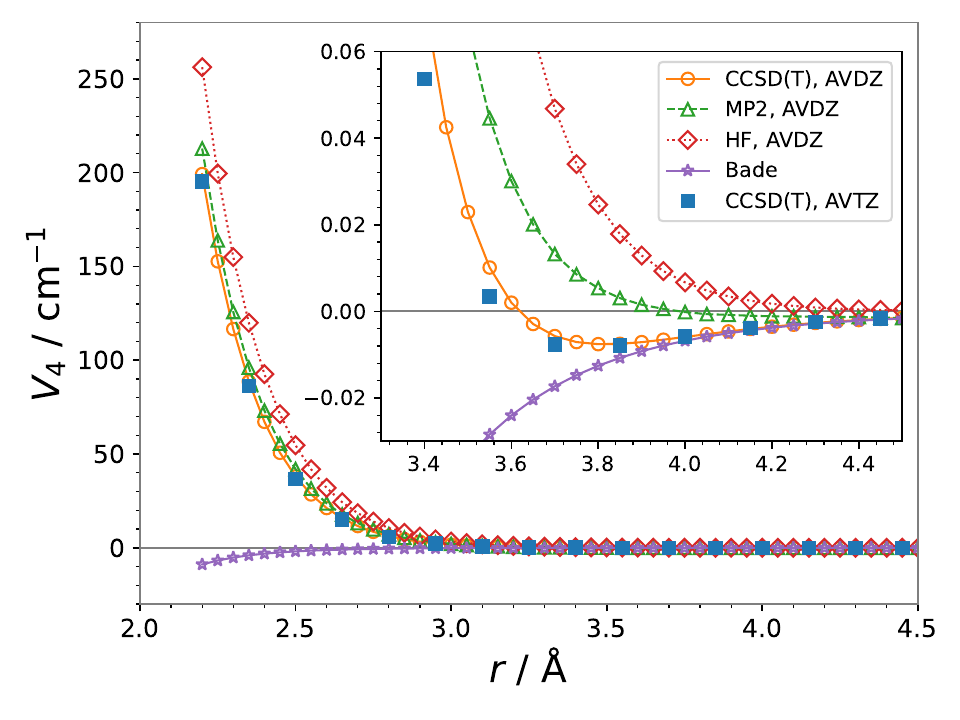}
    \caption{
        The four-body interaction potential energies of a tetrahedron,
        for side lengths between $ 2.2 \ang $ and $ 4.5 \ang $.
        Shown are the energies calculated for \{theory level, basis set\} combinations of
        \{CCSD(T), AVDZ\} (orange circles),
        \{CCSD(T), AVTZ\} (solid blue squares),
        \{MP2, AVDZ\} (green triangles), and
        \{HF, AVDZ\} (red diamonds),
        alongside the Bade potential (purple stars).
        We use the Midzuno-Kihara-type estimate of the Bade coefficient in the Bade potential
        shown here, although the difference between that estimate and the other estimates isn't
        visible at this scale.
    }
    \label{fig:paper3_method_comparison}
\end{figure}

The results are shown in Fig.~\ref{fig:paper3_method_comparison}.
We see that the CCSD(T) energies calculated using
both the AVDZ and AVTZ basis sets
are qualitatively and quantitatively very similar.
The MP2 curve is qualitatively different from both CCSD(T) curves;
it is more repulsive overall,
and does not have the same ``dip'' around $ 3.8 \ang $.
At short ranges,
the MP2/AVDZ and CCSD(T)/AVDZ energies differ from the CCSD(T)/AVTZ energies
by around $ 10 \, \% $ and $ 2 \, \% $, respectively.
The HF curve is the most repulsive of all the curves,
and shows no disperion interaction whatsoever at long ranges.
The analytic dispersive Bade potential is attractive for the entire range,
and also much weaker.
At long distances,
the CCSD(T)/AVDZ and CCSD(T)/AVTZ curves converge to the Bade potential,
which is a good indication of their quality.

\begin{figure} [ht]
    \centering
    \includegraphics[width=\linewidth]{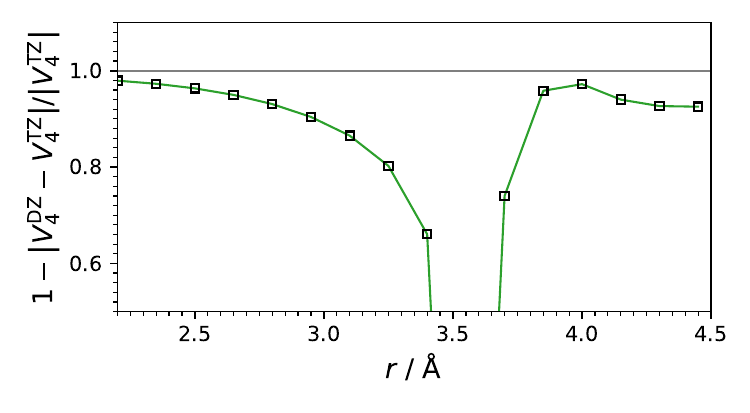}
    \caption{
    The relative error
    of calculating the four-body interaction energy
    of a tetrahedron
    with the AVDZ basis set relative to the AVTZ basis set
    as a function of the tetrahedron's side length $ r $.
    The dip in the graph occurs around where
    the four-body energy crosses $ 0 $.
    }
    \label{fig:paper3_avdz_avtz_relative_error}
\end{figure}

In Fig.~\ref{fig:paper3_avdz_avtz_relative_error},
we show the relative error of calculating the four-body interaction energy
of a tetrahedron with an AVDZ basis set relative to an AVTZ basis set.
The two curves are more similar in form
at short distances and at long distances.
The relative error naturally worsens around $ 3.6 \ang $,
which is roughly where the interaction energies cross $ 0 $,
as can be seen in Fig.~\ref{fig:paper3_method_comparison}.

Because basis sets are finitely sized,
all our calculations experience a systematic basis set size error (BSSE).
One way to estimate the BSSE
is to perform the same calculation with basis sets of several sizes,
and extrapolate the energies to that of an infinitely large basis.
We select two common extrapolation functions,
the exponential decay function,
given by
\begin{equation} \label{eq:basis_set_and_method:exponential}
    E_{e}(N) =
        E^{(\infty)}_{e} + c_{e} \exp\{ - \alpha N \}
\end{equation}
\noindent
and the power decay function,
given by
\begin{equation} \label{eq:basis_set_and_method:power}
    E_{p}(N) =
        E^{(\infty)}_{p} + c_{p} \, N^{-\beta} \, .
\end{equation}
In both equations,
$ N $ is the basis set cardinality (AVDZ = 2, AVTZ = 3, AVQZ = 4, etc.),
$ E^{(\infty)}_{e} $ and $ E^{(\infty)}_{p} $
are the energies for an infinitely large basis set
as predicted by the exponential and power decay fits respectively,
and $ c_{e} $, $ c_{p} $, $ \alpha $, and $ \beta $ are constants.

\begin{figure} [ht]
    \centering
    \includegraphics[width=\linewidth]{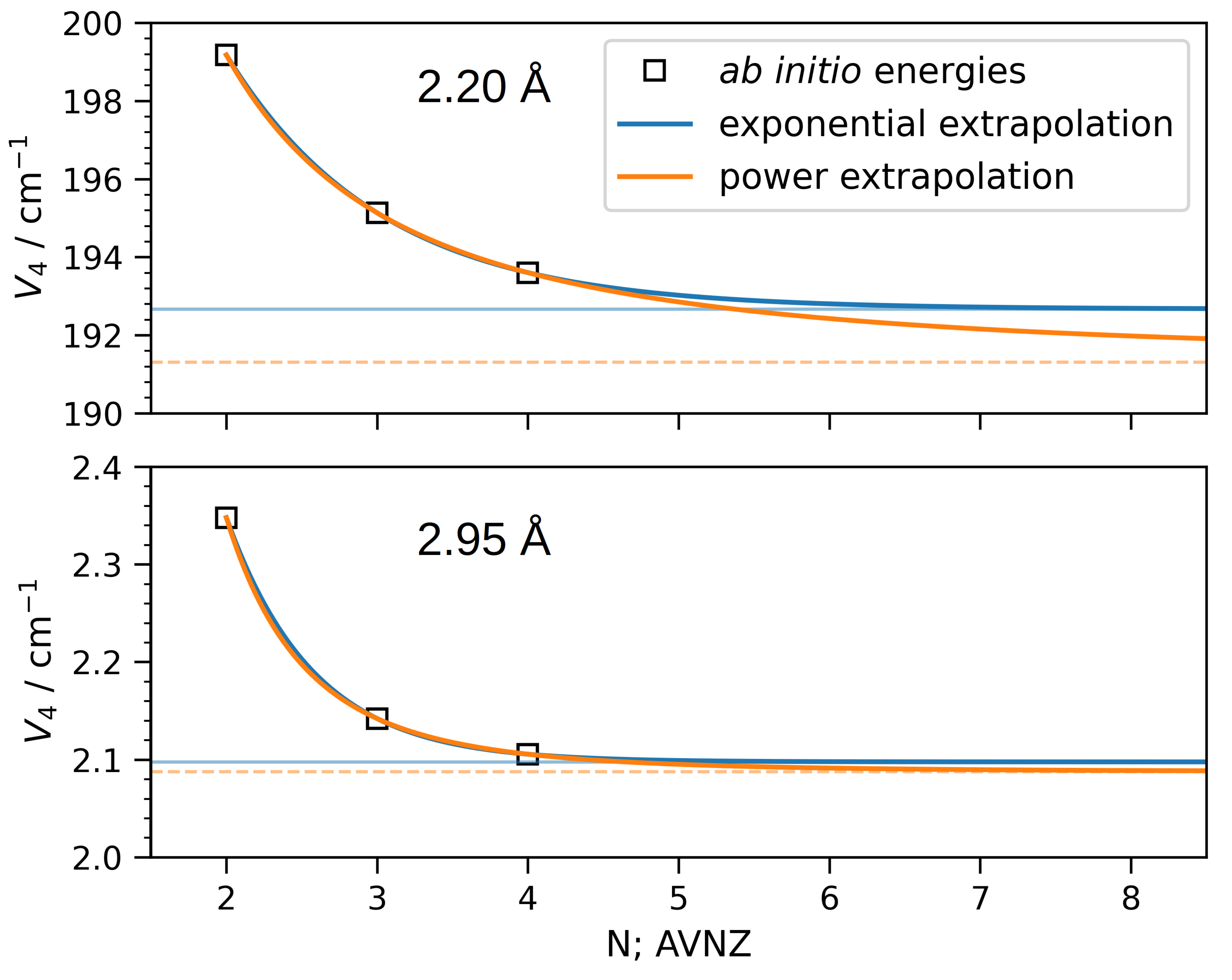}
    \caption{
        The four-body interaction energy of a tetrahedron
        of side length $ 2.2 \ang $ (top) and
        of side length $ 2.2 \ang $ (bottom),
        as a function of the basis set size (AVDZ = 2, AVTZ = 3, AVQZ = 4, etc.).
        The energies are extrapolated
        using the exponential decay extrapolation (blue curve)
        and the power decay extrapolation (orange curve).
        The horizontal lines represent the energies
        when extrapolated to a basis of infinite size.
    }
    \label{fig:basis_set_extrapolation}
\end{figure}

In Fig.~\ref{fig:basis_set_extrapolation},
we show the four-body interaction energy
of tetrahedra of side lengths $ 2.2 \ang $ and $ 2.95 \ang $
as a function of the basis set cardinality,
using both extrapolation functions.
The fit constants are provided in the supplementary material.
For a tetrahedron of side length $ 2.2 \ang $ ($ 2.95 \ang $),
the AVDZ energy captures
$ 97.9 \, \% $ ($ 90.4 \, \% $) of the AVTZ energy,
$ 97.1 \, \% $ ($ 88.5 \, \% $) of the AVQZ energy,
$ 96.6 \, \% $ ($ 88.1 \, \% $) of the exponentially extrapolated energy, and
$ 95.9 \, \% $ ($ 87.5 \, \% $) of the power decay extrapolated energy.
The power decay function gives a much more pessimistic
prediction of the convergence rate
than the exponential decay function
for the smaller tetrahedron.

The BSSE from using the AVDZ basis set is greater for larger geometries.
Similarly,
Wheatley \etal found that
the difference between AVTZ and AVQZ calculations for four-body systems of helium,
relative to the energy,
increases with the size of the geometry.\cite{fourbody:23whea}
The basis set size error
from using an AVDZ basis set
exceeds the fit error from the neural network
for the $ 64\sh128\sh128\sh64 $ and $ 64\sh128\sh128\sh64\sh\text{SSP} $ models.
We should also note that,
compared to the AVDZ calculations,
the AVTZ and AVQZ calculations respectively take
an average of 25 and 420 times as long,
and require about 4 and 30 times as much memory.

To put these errors into perspective,
one important use case for this PES is in
numerical simulations of \parahyd molecules
alongside the FSH pair potential and the recent three-body PES.
The FSH potential predicts a pair interaction energy
of $ 1079.4 \wvn $ between two \parahyd molecules $ 2.2 \ang $ apart.
The recent three-body PES predicts a three-body interaction energy
of $ -578.4 \wvn $ between three \parahyd molecules in an equilateral triangle of side length $ 2.2 \ang $.
The four-body interaction energy for a tetrahedron
with a side length of $ 2.2 \ang $ is roughly $ 200 \wvn $.
At $ 2.95 \ang $,
the two-body, three-body, and four-body energies for the aforementioned shapes
are $ 19.0 \wvn $, $ -16.4 \wvn $, and about $ 2.1 \wvn $, respectively.
Keeping in mind that
the four-body interaction is itself typically viewed as a correction term
to the two-body and three-body interactions,
the BSSE for this PES is probably tolerable for most applications,
even at larger geometries.

\subsection{Short Range: Exponential Decay} \label{sec:short_range}
We have no training samples
with a side length less than $ 2.2 \ang $,
and thus we need a method to extrapolate our energies to smaller geometries.

Consider the function $ V_4( \{ s r_{ij} \} ) $,
where $ \{ r_{ij} \} $ denotes the six side lengths
with the condition that the shortest among them is $ 2.2 \ang $,
and $ s \approx 1 $ is a scalar.
By modifying $ s $,
we scale all the side lengths by the same factor.
The four-body \parahyd interaction energy $ V_4( \{ s r_{ij} \} ) $
generally scales exponentially as a function of $ s $.
We can see that this is the case for the tetrahedron geometry
in Fig.~\ref{fig:paper3_method_comparison}.
In fact,
it is the case for all $ 8 $ of the highest energy \textit{hcp} geometries
shown in Fig.~\ref{fig:solid_parahydrogen:paper3_hcp_geometry_contribution_to_total_energy},
and the majority of the lower-energy \textit{hcp} geometries as well.
Only some of the \textit{hcp} geometries with very weak energies
do not fit well as a function of $ s $ to the exponential decay function.
However,
because they are weak,
the error from this poor fit is generally insignificant as long as the
energies aren't extrapolated to extremely short distances.
We use an exponential decay fit to extrapolate
the four-body \parahyd interaction PES
to geometries where the shortest side length is less than $ 2.2 \ang $.
The details are provided in the supplementary material.

\subsection{Long Range: Bade Potential} \label{sec:bade_potential}
The quadruple-dipole dispersion potential\cite{fourbody:57bade, fourbody:58bade}
is
\begin{multline} \label{eq:bade_potential:bade_potential_split}
V_{\rm B}( \{ \vb{r}_i \} )
=
V^{(2)}_{\rm B}( \{ \vb{r}_i \} )
+ V^{(3)}_{\rm B}( \{ \vb{r}_i \} )
+ V^{(4)}_{\rm B}( \{ \vb{r}_i \} )
\end{multline}
\noindent
where the ``pair component'' is given by
\begin{equation} \label{eq:bade_potential:pair_component}
V^{(2)}_{\rm B}( \{ \vb{r}_i \} )
=
- B_{\rm 12} \sum_{i<j}^{4}   r_{ij}^{-12} ,
\end{equation}
\noindent
the ``triplet component'' is given by
\begin{equation} \label{eq:bade_potential:triplet_component}
V^{(3)}_{\rm B}( \{ \vb{r}_i \} )
=
- B_{\rm 12}
\sum_{j=1}^{4} \sum_{i \neq j}^{3} \sum_{k > i, k \neq j}^{4}
\frac{(1 + (\vb{u}_{ij} \cdot \vb{u}_{jk}))^2}{r_{ij}^{6} r_{jk}^{6}} ,
\end{equation}
\noindent
and the ``quadruplet component'' is given by
\begin{equation} \label{eq:bade_potential:quadruplet_component}
V^{(4)}_{\rm B}( \{ \vb{r}_i \} ) = - 2 B_{\rm 12} [ f(1234) + f(1243) + f(1324) ] ,
\end{equation}
\noindent
where
\begin{multline}
f(ijkl) =
(r_{ij} r_{jk} r_{kl} r_{li})^{-3}
\big[
    -1 \\
    + (\vb{u}_{ij} \cdot \vb{u}_{jk})^2
    + (\vb{u}_{ij} \cdot \vb{u}_{kl})^2
    + (\vb{u}_{ij} \cdot \vb{u}_{li})^2 \\
    + (\vb{u}_{jk} \cdot \vb{u}_{kl})^2
    + (\vb{u}_{jk} \cdot \vb{u}_{li})^2
    + (\vb{u}_{kl} \cdot \vb{u}_{li})^2 \\
    - 3 (\vb{u}_{ij} \cdot \vb{u}_{jk}) (\vb{u}_{jk} \cdot \vb{u}_{kl}) (\vb{u}_{kl} \cdot \vb{u}_{ij}) \\
    - 3 (\vb{u}_{ij} \cdot \vb{u}_{jk}) (\vb{u}_{jk} \cdot \vb{u}_{li}) (\vb{u}_{li} \cdot \vb{u}_{ij}) \\
    - 3 (\vb{u}_{ij} \cdot \vb{u}_{kl}) (\vb{u}_{kl} \cdot \vb{u}_{li}) (\vb{u}_{li} \cdot \vb{u}_{ij}) \\
    - 3 (\vb{u}_{jk} \cdot \vb{u}_{kl}) (\vb{u}_{kl} \cdot \vb{u}_{li}) (\vb{u}_{li} \cdot \vb{u}_{jk}) \\
    + 9 (\vb{u}_{ij} \cdot \vb{u}_{jk}) (\vb{u}_{jk} \cdot \vb{u}_{kl}) (\vb{u}_{kl} \cdot \vb{u}_{li}) (\vb{u}_{li} \cdot \vb{u}_{ij})
\big] .
\end{multline}
\noindent
In the above equations,
$ B_{12} $ is a constant,
and $ r_{ij} $ and $ \vb{u}_{ij} $
are the distance and unit vector from position $ \vb{r}_i $ to $ \vb{r}_j $, respectively.

To create the long-range correction to our PES,
we are only interested in the quadruplet component.
This is because when calculating
the four-body interaction energy in Eq.~(\ref{eq:energy_calculations:interaction_energy}),
the pair and triplet components are already subtracted out.
Thus, our analytic long-range interaction will only be Eq.~(\ref{eq:bade_potential:quadruplet_component}),
which we refer to as the Bade potential
in the remainder of the paper.

At long distances,
the Bade potential decays as $ r^{-12} $.
We plot the Bade potential alongside the \abinitio energies
in Fig.~\ref{fig:paper3_method_comparison}
for tetrahedra of side lengths between $ 2.2 \ang $ and $ 4.5 \ang $.
We see that the CCSD(T) curves and the Bade potential converge at long distances.
Unlike the \abinitio energies,
the Bade potential is net attractive and much weaker at short range.
This discrepancy mirrors the case for the three-body interactions,
wherein the ATM potential predicts the short-range energies to be weak and net repulsive,
whereas the three-body \abinitio energies are stronger and attractive.

\subsubsection*{Estimates of the $ B_{12} $ Coefficient}
There is much less information in the literature on
the interaction coefficient for the four-body dispersion interaction $ B_{12} $
than the $ C_6 $ coefficient for the pair interaction,
or even the $ C_9 $ coefficient for the three-body interaction.
We can approximate the $ C_6 $, $ C_9 $, and $ B_{12} $ using\cite{fourbody:57bade}
\begin{align}
    C_{6} &\approx  \frac{3}{4}   \hbar \omega_0 \alpha^2 \\
    C_{9} &\approx  \frac{9}{16}  \hbar \omega_0 \alpha^3 \\
    B_{12} &\approx \frac{45}{64} \hbar \omega_0 \alpha^4 \, ,
\end{align}
\noindent
where $ \alpha $ is the polarizibility of molecular hydrogen,
and $ \hbar \omega_0 $ is a coupling constant.
Combining these equations gives
\begin{equation} \label{eq:bade_potential:mk_inspired_coeff}
    B_{12} = \frac{5}{3} \frac{C_9^2}{C_6} \, .
\end{equation}
This relationship is similar to the Midzuno-Kihara approximation,\cite{threebody:56midz}
which relates $ C_9 $ and $ C_6 $ though $ C_9 / C_6 = 3 \alpha / 4 $.
Using values of
$ C_6 = 58203.6 \wvn \ang^6 $ from Ref.~[\onlinecite{ph2cluster:15schm}], and
$ C_9 = 34336.2 \wvn \ang^9 $ from Ref.~[\onlinecite{h2pes:08hind}],
we find an estimate of
$ B_{12} = 33760.1 \wvn \ang^{12} $.
We can also approximate $ B_{12} $ directly from the \abinitio CCSD(T) interaction energies
for the tetrahedron geometry.
The estimates of $ B_{12} $
from energies calculated with the AVDZ and AVTZ basis sets,
are $ 31667.0 \wvn \ang^{12} $ and $ 29492.8 \wvn \ang^{12} $,
respectively (see the supplementary material).
The Midzuno-Kihara-like estimate of $ B_{12} $
is in surprisingly good agreement with the estimates from the \abinitio energies,
given that it is derived from a combination of three approximations.

For the PES published with this paper,
we decide to use the AVTZ-based estimate for $ B_{12} $.
We should emphasize that the neural network presented with this paper
does not depend on the chosen value of $ B_{12} $.
Instead,
the PES adjusts the interaction energy of a sample at long ranges
either after or instead of evaluating the neural network.
For cases where the long-range accuracy is especially important,
the value of $ B_{12} $ can be easily replaced in the code
without having to retrain the model.

\subsubsection*{Transitions Between Different Parts of the PES} \label{sec:transitions_short_long}
With no intervention,
there are small discontinuities between the short-range, long-range, and multilayer perceptron
parts of the PES.
These discontinuities are typically very small,
and in situations where only energies are required,
are probably not too concerning.

However,
there are many situations where it would be desirable to calculate forces with the PES.
The Python functor for the PES
provided in the accompanying repository
wraps around the different parts of the PES
introduces continuous transitions between them.
When used with the $ 64\sh128\sh128\sh64\sh\text{SSP} $ model,
this PES is continuously differentiable.
The conceptual details of the transition between the short-range, long-range, and multilayer perceptron
parts of the PES are provided in the supplementary material.

\subsection{Interactions in Solid Parahydrogen} \label{sec:solid_parahydrogen}
Solid \parahyd is a hexagonal close-packed (\textit{hcp}) crystal.\cite{ph2solidexp:80silv}
Even though solid \parahyd is a quantum solid
whose molecules are very delocalized about their nominal lattice sites,
we can gain several insights from analyzing its classical frozen lattice configuration.

Earlier,
we mentioned generating training data based on the geometries of the \textit{hcp} lattice.
We select a reference molecule in the lattice,
and generate all four-body geometries involving this reference molecule
such that
(1) at least one side length is equal to the lattice constant, and
(2) no side length is more than twice the lattice constant.
Geometries that do not obey these conditions have a nearly negligible contribution
to the total interaction energy,
even at a lattice constant as low as $ 2.2 \ang $.
There are $ 83 $ distinct four-body shapes that satisfy these conditions,
each occurring with a certain frequency.
A subset of the geometries are presented
in Table~\ref{tab:solid_parahydrogen:geometry_contributions}.
It shows the six side lengths $ \{ r_{ij} \} $ of the geometry normalized by the lattice constant,
and the number of times it appears subject to the aforementioned conditions $ N_c $.
The geometries are listed in order of increasing average side length.
The full table with all $ 83 $ geometries,
including sample energy values
is provided in the supplementary material.

\begin{figure*}
    \centering
    \includegraphics[width=\linewidth]{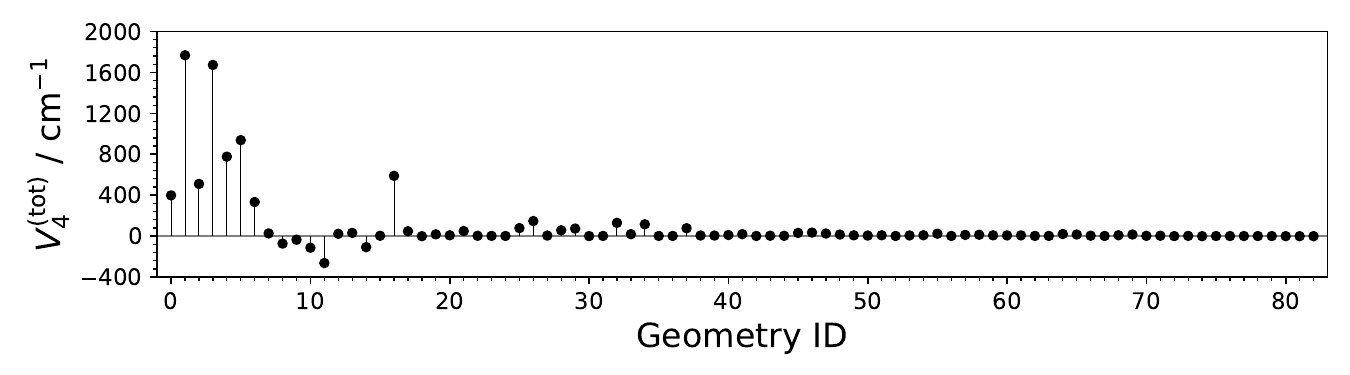}
    \caption{
        The total contribution of each geometry
        to the average four-body \parahyd interaction potential energy per particle
        in an \textit{hcp} lattice,
        when the lattice constant is $ 2.2 \ang $ ($ \rho \approx 0.133 \icubang $).
        The corresponding geometry IDs are provided in the supplementary material.
        The geometries are labelled in order of increasing average side length.
    }
    \label{fig:solid_parahydrogen:paper3_hcp_geometry_contribution_to_total_energy}
\end{figure*}

Consider a classical frozen \textit{hcp} lattice with a lattice constant of $ 2.2 \ang $.
For each geometry,
we take its four-body \parahyd interaction energy $ V_4(\{ r_{ij} \}) $,
multiply it with the number of times that geometry appears $ N_c $,
and divide by $ 4 $ to account for multiple counting.
This gives us $ V_4^{\rm (tot)} $,
the total contribution of each geometry
to the average four-body \parahyd interaction potential energy per particle
in the \textit{hcp} lattice.
In Fig.~\ref{fig:solid_parahydrogen:paper3_hcp_geometry_contribution_to_total_energy},
we show $ V_4^{\rm (tot)} $ for each geometry ID
in Table~\ref{tab:solid_parahydrogen:geometry_contributions}.

We see that nearly all the geometries provide a repulsive contribution
for the four-body \parahyd interaction energy.
We also see that the majority of geometries have nearly no contribution.
Of the 8 geometries with a total contribution greater than $ 300 \wvn $,
7 of them have the lowest geometry ID numbers
(and thus the shortest average side lengths).
However,
the average pair distance itself cannot explain the average interaction energy on its own.
Consider geometries 15 and 16
in Table~\ref{tab:solid_parahydrogen:geometry_contributions}.
These two geometries
have the same six side lengths but different physical structures
(none of the $ 24 $ allowed index permutations turn either geometry into the other).
At a lattice constant of $ 2.2 \ang $,
they have drastically different total contributions $ V_4^{\rm (tot)} $
of $ 3.99 \wvn $ and $ 589.18 \wvn $, respectively.

The four-body interaction potential for \parahyd
shares many features with the four-body interaction potential of helium.\cite{fourbody:23whea}
Both are repulsive at short range.
At larger intermolecular separations,
the energies are a mixed of repulsive and attractive,
although the repulsive energies tend to be stronger
and cover a larger amount of the coordinate space.

\begin{table} [h]
    \caption{
        The geometries of the \parahyd molecules in an \textit{hcp} lattice.
        The first column shows the geometry ID label.
        The geometries are labelled in order of increasing side length.
        The second column shows $ N_c $,
        the number of times that geometry appears in the first two shells.
        The remaining six columns show the side lengths of the geometries,
        normalized by the lattice constant.
        This table only shows 17 of the 83 geometries that satisfy the
        conditions described in the text.
        A full table is provided in the supplementary material.
    }
    \label{tab:solid_parahydrogen:geometry_contributions}
    \begin{tabular}{cccccccc}
            \hline
        \hline
        \mbox{ID} & $ N_c $ & $ r_{12} $ & $ r_{13} $ & $ r_{14} $ & $ r_{23} $ & $ r_{24} $ & $ r_{34} $ \\
        \hline
        $  0 $ & $   8 $ & $                    1 $ & $                    1 $ & $                    1 $ & $                    1 $ & $                    1 $ & $                    1 $  \\
        $  1 $ & $  48 $ & $                    1 $ & $                    1 $ & $                    1 $ & $                    1 $ & $                    1 $ & $             \sqrt{2} $  \\
        $  2 $ & $  12 $ & $                    1 $ & $                    1 $ & $                    1 $ & $                    1 $ & $                    1 $ & $  \sqrt{\sfrac{8}{3}} $  \\
        $  3 $ & $  36 $ & $                    1 $ & $                    1 $ & $                    1 $ & $                    1 $ & $                    1 $ & $             \sqrt{3} $  \\
        $  4 $ & $  12 $ & $                    1 $ & $                    1 $ & $             \sqrt{2} $ & $             \sqrt{2} $ & $                    1 $ & $                    1 $  \\
        $  5 $ & $ 144 $ & $                    1 $ & $                    1 $ & $                    1 $ & $                    1 $ & $             \sqrt{2} $ & $             \sqrt{3} $  \\
        $  6 $ & $  48 $ & $                    1 $ & $                    1 $ & $                    1 $ & $                    1 $ & $             \sqrt{2} $ & $ \sqrt{\sfrac{11}{3}} $  \\
        $  7 $ & $  24 $ & $                    1 $ & $                    1 $ & $                    1 $ & $             \sqrt{2} $ & $             \sqrt{2} $ & $  \sqrt{\sfrac{8}{3}} $  \\
        $  8 $ & $  72 $ & $                    1 $ & $                    1 $ & $                    1 $ & $                    1 $ & $             \sqrt{3} $ & $             \sqrt{3} $  \\
        $  9 $ & $  48 $ & $                    1 $ & $                    1 $ & $                    1 $ & $                    1 $ & $  \sqrt{\sfrac{8}{3}} $ & $ \sqrt{\sfrac{11}{3}} $  \\
        $ 10 $ & $  48 $ & $                    1 $ & $                    1 $ & $                    1 $ & $                    1 $ & $             \sqrt{3} $ & $ \sqrt{\sfrac{11}{3}} $  \\
        $ 11 $ & $  96 $ & $                    1 $ & $                    1 $ & $                    1 $ & $                    1 $ & $             \sqrt{3} $ & $                    2 $  \\
        $ 12 $ & $  24 $ & $                    1 $ & $                    1 $ & $             \sqrt{2} $ & $             \sqrt{2} $ & $                    1 $ & $ \sqrt{\sfrac{11}{3}} $  \\
        $ 13 $ & $  24 $ & $                    1 $ & $                    1 $ & $                    1 $ & $             \sqrt{2} $ & $             \sqrt{2} $ & $                    2 $  \\
        $ 14 $ & $  24 $ & $                    1 $ & $                    1 $ & $                    1 $ & $                    1 $ & $ \sqrt{\sfrac{11}{3}} $ & $ \sqrt{\sfrac{11}{3}} $  \\
        $ 15 $ & $  48 $ & $                    1 $ & $                    1 $ & $                    1 $ & $             \sqrt{2} $ & $             \sqrt{3} $ & $             \sqrt{3} $  \\
        $ 16 $ & $ 144 $ & $                    1 $ & $                    1 $ & $             \sqrt{2} $ & $             \sqrt{3} $ & $                    1 $ & $             \sqrt{3} $  \\
        \hline
        \hline
    \end{tabular}
\end{table}

We can look at the impact of many-body interactions in rare gas solids,
such as solid helium-4, neon, argon, krypton, and xenon,
for insights on how they behave in solid \parahyd.
Like solid \parahyd,
the rare gas solids are molecular solids with weak intermolecular dispersion interactions,
and are common subjects in quantum path integral simulations.\cite{pathinteg:14herr}

There appears to be a common pattern in many-body interactions of rare gas atoms.
The overall interaction energy at short intermolecular separations
switches signs based on the parity of the interaction order.
At high densities in solid helium-4,
Tian \etal found that
the 2-body, 4-body, and 6-body interactions have net repulsive contributions,
while the 3-body and 5-body interactions have net attractive contributions.\cite{manybody:06tian}
The magnitude of the contributions also decrease with every increasing order.
Similarly, at high densities in solid krypton,
the 2-body and 4-body interactions have overall repulsive contributions,
while the 3-body interaction is attractive.\cite{fourbody:12tian}
In solid neon, argon, and xenon,
the three-body interaction energy is known to be overall attractive.\cite{threebody:00rosc}
It appears that many-body interactions between \parahyd
molecules also follow this pattern.

For the equation of state,
we calculate the energy per molecule $ \epsilon $ as a function of the density $ \rho $
for a solid \parahyd \textit{hcp} lattice with no zero-point motion.
We can then calculate the pressure $ P $ as a function of density using
\begin{equation} \label{eq:solid_parahydrogen:pressure_density_relation}
    P = \rho^2 \pdv{\epsilon}{\rho}\eval_{T}
\end{equation}
\noindent
For these calculations,
we use the FSH two-body potential,\cite{ph2cluster:14faru}
the recent isotropic three-body PES,\cite{threebody:22ibra}
and the 64-128-128-64 four-body PES presented in this paper.
    \begin{figure} [ht]
        \centering
    \includegraphics[width=\linewidth]{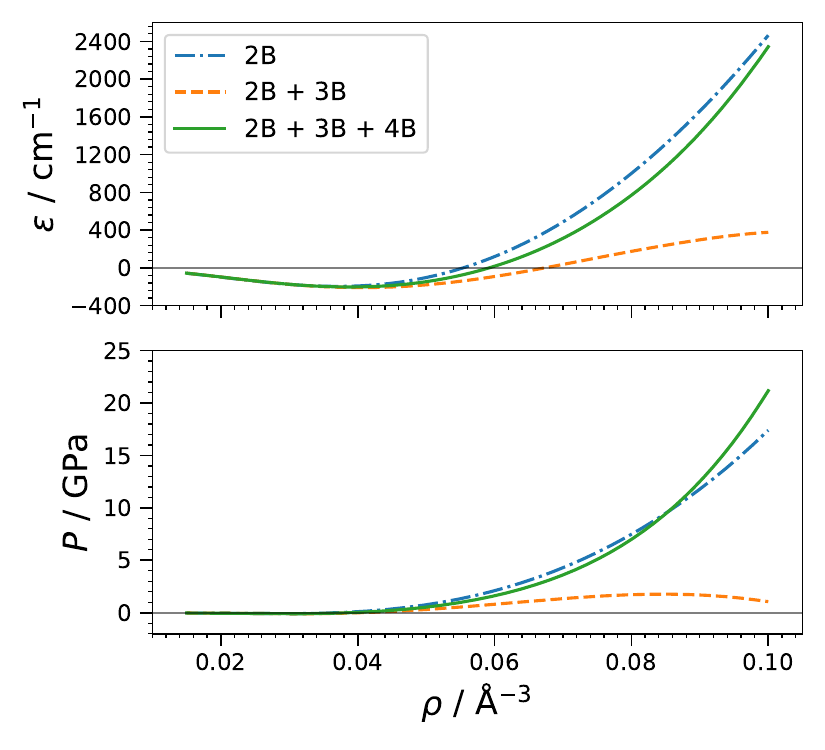}
    \caption{
    We show the classical energy-density curves (top)
    and classical pressure-density curves (bottom),
    for a frozen \textit{hcp} lattice whose interactions are made up of
    the two-body potential on its own (blue, dash-dotted),
    the two-body and three-body potentials together (orange, dashed),
    and the two-, three-, and four-body potentials together (green, solid).
    The top and bottom subfigures both share the same horizontal axis.
    The highest density shown here,
    $ \rho = 0.1 \icubang $,
    corresponds to a lattice constant of $ a = 2.42 \ang $.
    }
    \label{fig:solid_parahydrogen:paper3_classical_energy_and_pressure_density}
\end{figure}
In Fig.~\ref{fig:solid_parahydrogen:paper3_classical_energy_and_pressure_density},
we show the classical energy-density curves (top)
and classical pressure-density curves (bottom),
for a lattice whose interactions are made up of the two-body potential on its own,
the two-body and three-body potentials together,
and the two-, three-, and four-body potentials together.
We see that, below about $ 0.08 \icubang $,
the addition of the four-body interaction
moves the pressure-density curve
below the case where the two-body PES is used on its own,
and above the case where the two-body and three-body PESs are present.
Qualitatively,
these results are promising.
In an earlier study of PIMC simulations of solid parahydrogen,
we found that the pair potential overestimates the pressure-density curve
compared to experiment.\cite{pathinteg:19ibra}
However,
the inclusion of the three-body potential causes an overcorrection
at higher densities,\cite{pathinteg:22ibra}
causing the simulation to underestimate the pressure-density curve.
Except at very high densities,
the four-body potential should place the pressure-density curve
between these two former cases,
bringing it closer to experiment.

\section{Conclusion} \label{sec:conclusion}
We have presented an isotropic \abinitio four-body interaction PES for \parahyd.
The electronic structure calculations are performed at the CCSD(T) level
with an AVDZ atom-centred basis set,
supplemented by a $(3s3p2d)$ midbond function.
The energies are fit to a multilayer perceptron.
The procedure used to construct the PES has been provided in detail.
This includes the methods of sample generation,
the transformations on the input side lengths and the output energies
to help improve the training process,
and the training of the neural network itself.

We have also discussed how the interaction energy is extrapolated
outside the training data's regime,
using an analytic dispersion interaction at long intermolecular separations,
and an exponential decay when the \parahyd molecules are close together.
The four-body \parahyd interaction energy is overall repulsive at high densities.
To put this into context,
we know from previous PIMC simulations that
the \textit{ab initio} two-body FSH PES on its own is too repulsive
at high densities,\cite{pathinteg:19ibra}
and that the inclusion of the \textit{ab initio} three-body PES
is too attractive at high densities.\cite{pathinteg:22ibra}
Using a classical approximation,
we show that the additional inclusion of the four-body PES
will likely improve the agreement of the simulation results
with experiment,
except at very high densities.

\section*{SUPPLEMENTARY MATERIAL}
Refer to the supplementary material for detailed descriptions of the 
Estimates of Bade Coefficients, Lexicographic Ordering, 
Training Error Curves for Smaller Models, Short Range Exponential Decay Fit, 
and Contributions to the \textit{hcp} lattice for each geometry.

\section*{Acknowledgements}
The authors acknowledge the Natural Sciences and Engineering Research Council (NSERC) of Canada (RGPIN-2016-04403), the Ontario Ministry of Research and Innovation (MRI), the Canada Research Chair program (950-231024), and the Canada Foundation for Innovation (CFI) (project No. 35232). A.~I. acknowledges the support of the NSERC of Canada (CGSD3-558762-2021).

\section*{DATA AVAILABILITY}
The data that support the findings of this study are openly available at \url{https://github.com/AlexanderIbrahim1/nn_fourbody_potential/}.
The raw inputs and outputs involved in performing the electronic structure calcultions for the \textit{ab initio} energies are available at DOI \url{10.5281/zenodo.11272857}.

\section*{References}

\bibliography{biblio}
\bibliographystyle{ieeetr}

\end{document}



\title{Supplementary Material for: A neural network-based four-body potential energy surface for parahydrogen}

\author{Alexander Ibrahim}
\affiliation{
    Department of Physics and Astronomy,
    University of Waterloo,
    200 University Avenue West, Waterloo, Ontario N2L 3G1, Canada
}
\affiliation{
    Department of Chemistry,
    University of Waterloo,
    200 University Avenue West, Waterloo, Ontario N2L 3G1, Canada
}

\author{Pierre-Nicholas Roy}
\email{pnroy@uwaterloo.ca}
\affiliation{
    Department of Chemistry,
    University of Waterloo,
    200 University Avenue West, Waterloo, Ontario N2L 3G1, Canada
}

\maketitle

\section{Trilateration}
\red{In the sample generation procedure,
we sample the six side lengths of the geometry
\begin{equation*}
    (r_{12}, r_{13}, r_{14}, r_{23}, r_{24}, r_{34})
\end{equation*}
\noindent
and use a trilateration procedure to generate the Cartesian coordinates.}
\red{The function $ \texttt{SixSideLengthsToCartesian} $ below
accepts the six side lengths of the four-body geometry,
and a small number $ \varepsilon > 0 $ to help account for
a corner case involving the position of point 3 (described later).
If the six side lengths represent a valid four-body geometry,
the function returns four Cartesian coordinates that generate the geometry.
Otherwise,
the function returns \textbf{None}.}

\algnewcommand{\LineComment}[1]{\State \(\triangleright\) #1}
\begin{algorithmic}
    \Function{SixSideLengthsToCartesian}{$r_{12}, r_{13}, r_{14}, r_{23}, r_{24}, r_{34}, \varepsilon$}
        \LineComment{calculate the relevant $x$ positions}
        \State $ x_2 \gets r_{12} $
        \State $ x_3 \gets (r_{12}^2 + r_{13}^2 - r_{23}^2) / 2 r_{12} $
        \State $ x_4 \gets (r_{14}^2 - r_{24}^2 + r_{12}^2) / 2 r_{12} $

        \
        \LineComment{calculate the $y$ positions, accounting for a corner case and floating-point errors}

        \State $ \tilde{y_3}^2 \gets r_{13}^2 - x_3^2 $

        \
        \LineComment{indication that the geometry is impossible}
        \If{$ \tilde{y_3}^2 < 0 \,\, \textbf{and} \,\, |\tilde{y_3}^2| > \varepsilon $}
            \State \Return \textbf{None}
        \EndIf

        \

        \If{$ \tilde{y_3}^2 > \varepsilon $}
            \State $ y_3 \gets \sqrt{\tilde{y_3}^2} $
            \State $ y_4 \gets (r_{14}^2 - r_{34}^2 + r_{13}^2 - 2 x_3 x_4) / 2 y_3 $
        \Else
            \State $ y_3 \gets 0 $
            \State $ y_4 \gets 0 $
        \EndIf

        \
        \LineComment{calculate $z_4$, accounting for floating-point errors}

        \State $ \tilde{z_4}^2 \gets r_{14}^2 - x_4^2 - y_4^2 $

        \
        \LineComment{indication that the geometry is impossible}
        \If{$ \tilde{z_4}^2 < 0 \,\, \textbf{and} \,\, |\tilde{z_4}^2| > \varepsilon $}
            \State \Return \textbf{None}
        \EndIf

        \
        \State $ z_4 \gets \sqrt{\text{max}(0, \tilde{z_4}^2)} $

        \
        \LineComment{All coordinates needed to generate the points have been calculated}

        \State $ (p_1, p_2, p_3, p_4) \gets \{ (0, 0, 0), (x_2, 0, 0), (x_3, y_3, 0), (x_4, y_4, z_4) \} $
        \State \Return $ (p_1, p_2, p_3, p_4) $
    \EndFunction
\end{algorithmic}

\red{Using this procedure,
the generated points satisfy certain properties.
Point 1 is always at the origin.
Point 2 lies on the positive $x$-axis.
Point 3 lies in the region $ y \ge 0 $ and $ z = 0 $.
Point 4 lies in the region $ z \ge 0 $.}
\red{The two conditional statements that cause the function to return \textbf{None}
are those in which the six side lengths do not correspond to a valid geometry.
They indicate that the six side lengths violate the triangle inequality rule in
one or more ways.}
\red{In the general case where
point 3 lies in the $xy$-plane but not on the $x$-axis,
the procedure can assign point 4 a definite, unique position.
However,
if point 3 lies on the $x$-axis and $ y_3 = 0 $,
the calculation for $ y_4 $ blows up.
In this special case,
points 1, 2, and 3 all lie on the $x$-axis,
and point 4 may lie anywhere on a ring of radius $ \sqrt{r_{14}^2 - x_4^2} $
that lies parallel to the $yz$-place and is centred at $ x_4 $ on the $x$-axis.
We may arbitrarily select $ y_4 = 0 $ without modifying the relative pair distances
of the system.}
\red{Calculating $ y_3 $ and $ z_4 $ each involve taking a square root,
and in cases where $ y_3 $ and $ z_4 $ are very close to or essentially $ 0 $,
floating-point errors can cause the arguments to the square root to be negative.
The above function accounts for these cases.}

\section{Estimates of Bade Coefficients}

\begin{figure} [H]
    \centering
    \includegraphics[width=0.9\linewidth]{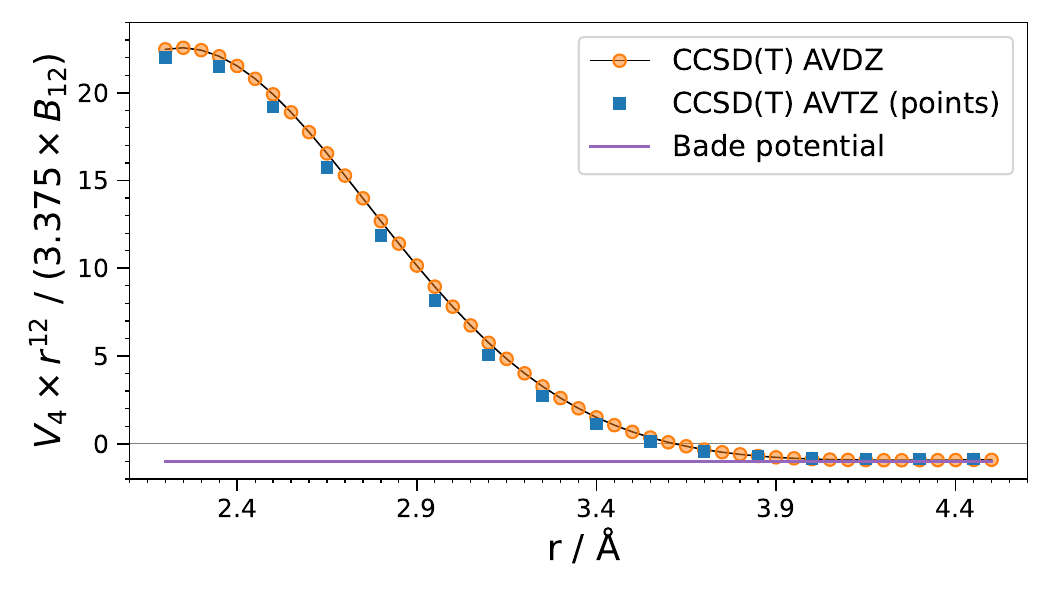}
    \includegraphics[width=0.9\linewidth]{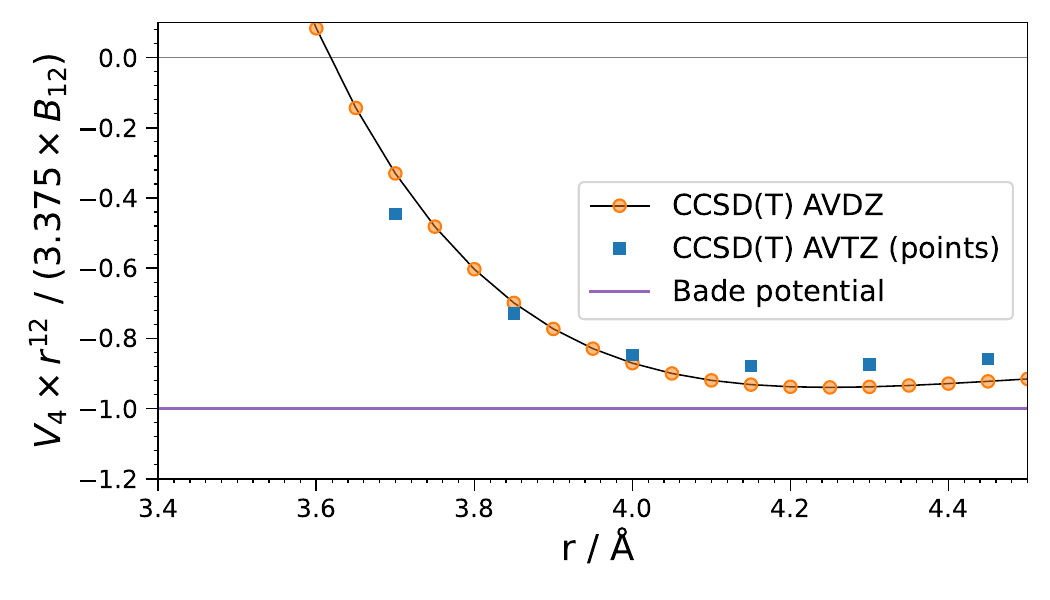}
    \caption{
        The rescaled four-body interaction energy for a tetrahedron geometry,
        as a function of the tetrahedron's side length $ r $.
        Shown are
        the CCSD(T) energies calculated
        using the AVDZ basis (orange circles, black line),
        and the AVTZ basis (blue squares),
        as well as the Bade potential (purple line).
        The term in the $y-$axis label, $ B_{12} $,
        is the Bade coefficient as estimated using the Midzuno-Kihara-inspired approximation.
        The top subfigure shows the results for the entire range of
        energies calculated for the tetrahedron,
        while the bottom subfigure emphasizes the region
        where the long-range $ r^{-12} $ decay begins.
    }
    \label{fig:bade_rescaled:bade_coefficient_estimation}
\end{figure}

In Fig.~\ref{fig:bade_rescaled:bade_coefficient_estimation},
we plot transformed versions of the Bade potential alongside
the \abinitio CCSD(T) AVDZ and AVTZ interaction energies for a tetrahedron
as a function of the tetrahedron side length $ r $.
For the transformations,
we multiply the interaction energies by $ r^{12} $,
and rescale all the curves by the same constant
such that the transformed Bade potential is equal to $ -1 $.
The figure makes it easier to see
that the \abinitio energies
have the expected $ r^{-12} $ decay rate at large intermolecular separations.
It is from where these rescaled curves level out
that we estimate the interaction coefficients $ B_{12} $
derived from the AVDZ and AVTZ \abinitio energies.

\section{Lexicographic Ordering}

Lexicographic ordering is a way to compare two ordered sequences based on their elements.
For readers familiar with the Python programming language,
this is the kind of ordering that Python uses
to compare two tuples of numbers. When comparing two sequences,
we compare elements at the same index in either sequence, from left to right.
If the two elements are equal,
we move on to compare the elements in the next index.
As soon as we hit an index where one element is lower than the other,
the sequence of the lower element is considered the lower sequence overall.
In terms of pseudocode, this can be represented as the following function \texttt{LexLowest},
which takes two sequences,
compares them lexicographically,
and returns the smallest of the two (or either if they are equal).

\begin{algorithmic}
    \Function{LexLowest}{$A, B$}
        \State $n \gets \text{length of } A$
        \For{$i \gets 1$ to $n$}
            \If{$A[i] < B[i]$}
                \State \Return $A$
            \ElsIf{$A[i] > B[i]$}
                \State \Return $B$
            \EndIf
        \EndFor
        \State \Return $A$   \text{
    \ \ \ \ \ \ \ \ \ \ \ \ \ \ \ \ \hspace{1em}$\blacktriangleright$ A and B are equal, return either
}
    \EndFunction
\end{algorithmic}

For example,
suppose we have input samples
\begin{align}
    s_0 &= (2, 3, 4, 5, 6, 7) \nonumber \\
    s_1 &= (2, 3, 4, 7, 1, 9) \nonumber \\
    s_2 &= (1, 9, 11, 25, 13, 15) \, .\nonumber
\end{align}
\noindent
When comparing these samples, we would find that
\texttt{LexLowest($s_0$, $s_1$)~=~$ s_0 $},
because of the comparison of $ 5 $ and $ 7 $ in the fourth position,
\texttt{LexLowest($s_0$, $s_2$)~=~$ s_2 $},
because of the comparison of $ 2 $ and $ 1 $ in the first position,
and \texttt{LexLowest($s_1$, $s_2$)~=~$ s_2 $},
again because of the comparison of $ 2 $ and $ 1 $ in the first position.

\section{Training Error Curves for Smaller Models}

\begin{figure} [H]
    \centering
    \includegraphics[width=0.7\linewidth]{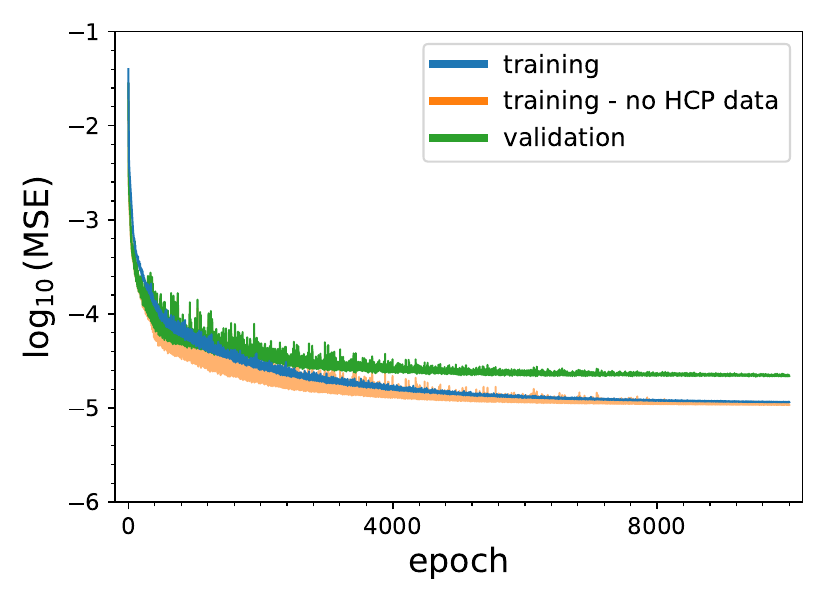}
    \caption{The base-10 logarithm of the mean squared error loss
    as a function of the epoch
    during the training of the $ 32\sh64\sh64\sh32 $ model.
    The plots are shown for the
    training data set (blue, solid),
    the validation data set (green, solid),
    and the training data set with the 1610 samples from
    the \textit{hcp} lattice removed (orange, semi-translucent).
    }
    \label{fig:neural_network:neural_network_error_32_64_64_32_far}
\end{figure}

\begin{figure} [H]
    \centering
    \includegraphics[width=0.7\linewidth]{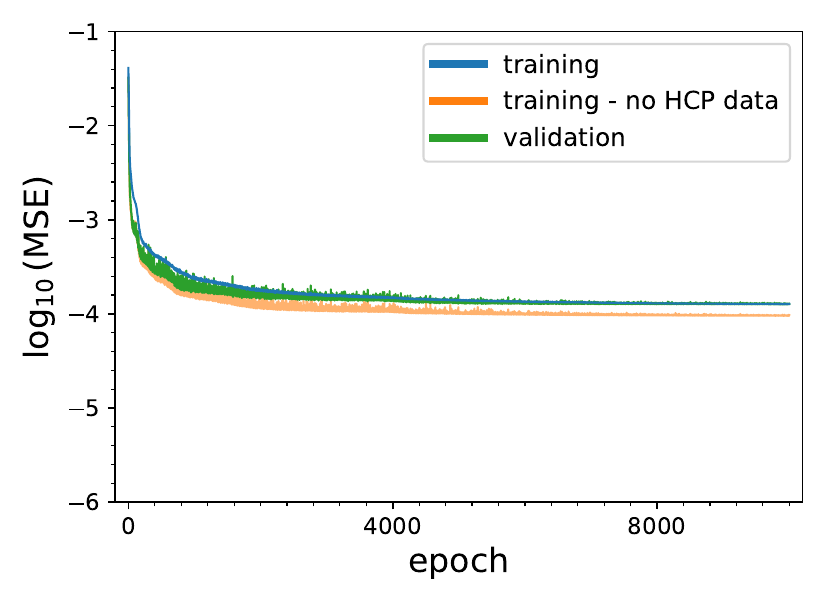}
    \caption{The base-10 logarithm of the mean squared error loss
    as a function of the epoch
    during the training of the $ 16\sh32\sh32\sh16 $ model.
    The plots are shown for the
    training data set (blue, solid),
    the validation data set (green, solid),
    and the training data set with the 1610 samples from
    the \textit{hcp} lattice removed (orange, semi-translucent).
    }
    \label{fig:neural_network:neural_network_error_16_32_32_16_far}
\end{figure}

\begin{figure} [H]
    \centering
    \includegraphics[width=0.7\linewidth]{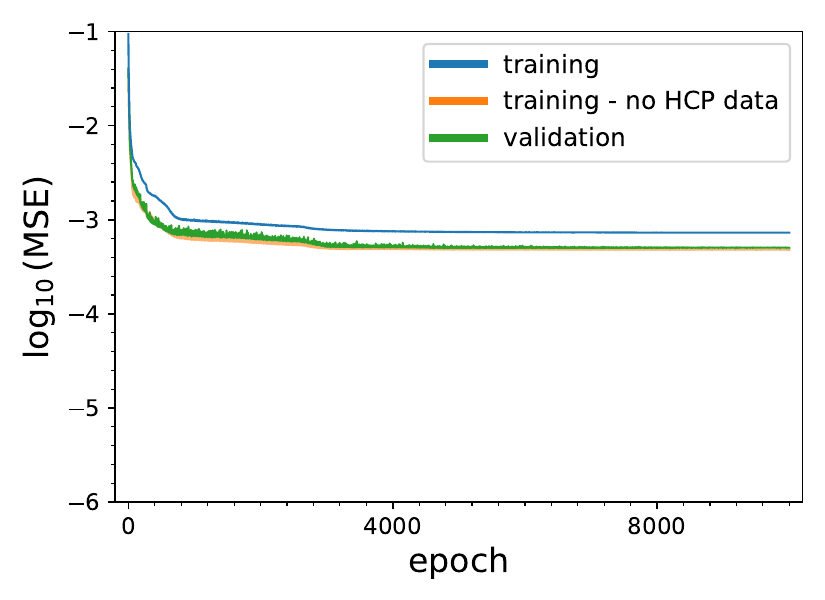}
    \caption{The base-10 logarithm of the mean squared error loss
    as a function of the epoch
    during the training of the $ 8\sh16\sh16\sh8 $ model.
    The plots are shown for the
    training data set (blue, solid),
    the validation data set (green, solid),
    and the training data set with the 1610 samples from
    the \textit{hcp} lattice removed (orange, semi-translucent).
    }
    \label{fig:neural_network:neural_network_error_8_16_16_8_far}
\end{figure}

The main text contains the training error of the neural network as a function of epoch,
for the $ 64\sh128\sh128\sh64 $ model.
Here we show the training curves for
the $ 32\sh64\sh64\sh32 $ model (Fig.~\ref{fig:neural_network:neural_network_error_32_64_64_32_far}),
the $ 16\sh32\sh32\sh16 $ model (Fig.~\ref{fig:neural_network:neural_network_error_16_32_32_16_far}), and
the $ 8\sh16\sh16\sh8 $ model (Fig.~\ref{fig:neural_network:neural_network_error_8_16_16_8_far}).

\section{Basis Set Superposition Error Fit Constants}
\red{In the Discussion and Analysis section of the main text,
in the subsection titled ``Basis Set and Method Comparison'',
we take the AVDZ, AVTZ, and AVQZ energies
for tetrahedra
and extrapolate them to infinite basis size using the
the exponential decay function}
\begin{equation} \label{eq:basis_set_and_method:exponential}
\red{
    E_{e}(N) =
        E^{(\infty)}_{e} + c_{e} \exp\{ - \alpha N \}
}
\end{equation}
\noindent
\red{and the power decay function}
\begin{equation} \label{eq:basis_set_and_method:power}
\red{
    E_{p}(N) =
        E^{(\infty)}_{p} + c_{p} \, N^{-\beta} \, .
}
\end{equation}

\red{For the tetrahedron of side length $ 2.2 \ang $,
the fit constants are
$ E^{(\infty)}_{e} = 192.668 \wvn $,
$ c_{e} = 45.341 \wvn $,
$ \alpha = 0.9707 $,
$ E^{(\infty)}_{p} = 191.311 \wvn $,
$ c_{p} = 26.999 \wvn $, and
$ \beta = 1.780 $.
For the tetrahedron of side length $ 2.95 \ang $,
the fit constants are
$ E^{(\infty)}_{e} = 2.0976 \wvn $,
$ c_{e} = 7.954 \wvn $,
$ \alpha = 1.729 $,
$ E^{(\infty)}_{p} = 2.0875 \wvn $,
$ c_{p} = 3.775 \wvn $, and
$ \beta = 3.858 $.}

\section{Short Range: Exponential Decay Fit}

For four-body geometries with at least one side length shorter than $ 2.2 \ang $,
we use an extrapolation function that is a linear combination of
an exponential decay fit and a linear fit.
For well-behaved cases,
the extrapolation function is entirely the exponential decay fit.
For cases where the exponential decay fit would behave poorly or blow up,
the linear fit becomes more prominent.

\red{Suppose we have a geometry $ g $,
made up of the six side lengths $ g = \{ r_{ij} \} $,
where at least one of the side lengths is less than $ 2.25 \ang $.
Define $ \Delta R = 0.01 \ang $ and $ \tilde{R}_{\rm U} = 2.25 \ang $,
where the subscript is chosen in the context of a later section in the supplementary material.
Let $ R_0 $ be the shortest side length in the input.
We calculate the scaling factors $ s_0 = \tilde{R}_{\rm U} / R_0 $ and $ s_1 = (\tilde{R}_{\rm U} + \Delta R) / R_0 $,}
and define two new geometries $ g_0 = \{ s_0 r_{ij} \} $ and $ g_1 = \{ s_1 r_{ij} \} $.

We now calculate the exponential coefficient
\begin{equation} \label{eq:supplementary:expon_coeff}
\red{
    c = \frac{1}{\Delta R} \ln \left| \frac{V_4(g_1)}{V_4(g_0)} \right| \, .
}
\end{equation}
\noindent
The energy for the original geometry $ g $ is now calculated using the exponential decay fit
\begin{equation} \label{eq:supplementary:expon_decay_fit}
\red{
    V_4^{\rm (ex)}(g) = V_4(g_0) \exp \{ -c (\tilde{R}_{\rm U} - R_0) \} \, .
}
\end{equation}
\noindent
However,
there are cases where $ c $ is very large,
and Eq.~(\ref{eq:supplementary:expon_decay_fit}) blows up very quickly.
On closer inspection,
this typically only happens in cases where $ V_4(g_0) $ is very close to 0,
which means a small absolute change in energy
results in a large relative change in energy.
To prevent this,
we smoothly transition from the exponential fit
to a less physically realistic, but more numerically stable linear fit.

The full equation is
\begin{equation} \label{eq:supplementary:full_extrapolation}
    V_4(g) = \omega(c; c', c'') V_4^{\rm (ex)}(g) + [1 - \omega(c; c', c'')] V_4^{\rm (li)}(g)
\end{equation}
\noindent
where $ V_4^{\rm (li)}(g) $ is a linear extrapolation below $ r = r_0 $,
$ c' = 6 $,
$ c'' = 8 $,
and \red{$ \omega(x; a, b) $ is given by}
\begin{equation} \label{eq:supplementary:smooth01_transition}
\red{
    \omega(x; a, b) = 
    \begin{cases}
        0                             & \text{if } x \leq a, \\
        \frac{1}{2} [1 - \cos(\pi k)] & \text{if } a < x < b, \\
        1                             & \text{if } x \geq b
    \end{cases}
}
\end{equation}
\noindent
\red{where $ k = (x - a)/(b - a) $}.
If $ c $ is below $ c' $,
the fit is entirely exponential;
if $ c $ is above $ c'' $,
the fit is entirely linear;
and if $ c $ is between them,
it is a finite linear combination of the two.
The constants $ c' $ and $ c'' $ are chosen by hand
after investigating certain cases where the exponential decay fit behaved poorly.
The linear fit purposely sacrifices accuracy for stability.
However,
\red{because the calculation of the exponential coefficient only blows up when the energies involved
are weak,}
the error from this linear adjustment is generally insignificant as long as the
energies aren't extrapolated to extremely short distances.
\red{The decision to set $ \tilde{R}_{\rm U} $ to $ 2.25 \ang $ rather than $ 2.2 \ang $
is done in anticipation of the use of a transition function described in the next section.}

\section{Transitions between different parts of the PES}
\subsection*{Description}

\red{The PES,
using the MLP trained with the shifted softplus activation function,
is continuously differentiable.}

\subsubsection*{Transition between \textit{ab initio} and dispersive parts}

\red{Conceptually,
the PES is split into an ``\textit{ab initio} part,'' a ``dispersive part,'' and a ``mixed (\textit{ab initio}/dispersive) part.''
The \textit{ab initio} part is made of the multilayer perceptron (MLP) and the exponential decay parts,
described in further detail later.
The dispersive part is made of the Bade potential.
The mixed (\textit{ab initio}/dispersive) part is made of a linear combination of the \textit{ab initio} and dispersive parts.}

\red{Each four-body geometry falls into one of these three categories
based on its average side length $ \tilde{r} $.
Define a ``lower'' distance $ \tilde{r}_{\rm L} $ and an ``upper'' distance $ \tilde{r}_{\rm U} $,
where $ \tilde{r}_{\rm L} < \tilde{r}_{\rm U} $.
The part of the PES that a sample falls into is described by}
\begin{equation}
\red{
    \text{part of PES} =
    \begin{cases}
        \textit{ab initio} & \text{if } \tilde{r} \leq \tilde{r}_{\rm L}, \\
        \text{mixed (\textit{ab initio}/dispersive)} & \text{if } \tilde{r}_{\rm L} < \tilde{r} < \tilde{r}_{\rm U}, \\
        \text{dispersive} & \text{if } \tilde{r} \geq \tilde{r}_{\rm U}
    \end{cases}
}
\end{equation}
\noindent
\red{In the mixed (\textit{ab initio}/dispersive) part,
the energy of a sample is a linear combination of the \textit{ab initio} and dispersive parts.
The fraction of the output that comes from the dispersive part is given by}
\begin{equation}
\red{
    \omega(\tilde{r}; \tilde{r}_{\rm L}, \tilde{r}_{\rm U}) = 
    \begin{cases}
        0                             & \text{if } \tilde{r} \leq \tilde{r}_{\rm L}, \\
        \frac{1}{2} [1 - \cos(\pi k)] & \text{if } \tilde{r}_{\rm L} < \tilde{r} < \tilde{r}_{\rm U}, \\
        1                             & \text{if } \tilde{r} \geq \tilde{r}_{\rm U}
    \end{cases}
}
\end{equation}
\noindent
\red{where $ k = (\tilde{r} - \tilde{r}_{\rm L}) / (\tilde{r}_{\rm U} - \tilde{r}_{\rm L}) $.
In the current PES,
$ \tilde{r}_{\rm L} = 4 \ang $ and $ \tilde{r}_{\rm U} = 4.5 \ang $.}

\red{The decision to make the transition between the \textit{ab initio} and dispersive parts
depend on the average side length,
and the choice of $ \tilde{r}_{\rm L} $ and $ \tilde{r}_{\rm U} $,
were based on observations of the \textit{hcp} lattice geometries.
By looking at the energies of geometries in the \textit{hcp} lattice
as a function of the lattice constant,
it appeared that when their average side length was around $ 4 \ang $,
the energy was sufficiently converged to the Bade potential.}

\subsubsection*{Transition within the \textit{ab initio} part}

\red{The \textit{ab initio} part is further split into a part where
the energies are determined directly by the MLP (the ``mid-range part''),
by the exponential decay extrapolation (the ``short-range part''),
and by a linear combination of the two (the ``mixed (short/mid)-range part'').}

\red{The exponential decay fit is described earlier in the supplementary material.
With no mixed (short/mid)-range part to transition between the two,
there is a small discontinuity between the mid-range and the short-range parts.
Depending on how small we make the ``scaling step size'' ($ \Delta R $) for the exponential decay fit,
we can make the discontinuity very small.
For example,
setting $ \Delta R = 0.01 \ang $
introduces a discontinuity at the short-range/mid-range transition
on the order of $ 0.1 \% $ for geometries with large energies
(such as the tetrahedron, and other ``compact'' geometries),
and on the order of $ 1-3 \% $ for geometries with much smaller energies
(on the order of $ 1 \wvn $).}
\red{For situations where only the energies are needed,
this discontinuity is likely unimportant.
For situations where the PES must be continuously differentiable,
we can remove this discontinuity by using the same transition function and strategy
we used for the transition between the \textit{ab initio} and dispersive regions.}

\red{A four-body geometry with an average side length less than $ \tilde{r}_{\rm U} $
falls into either the short-range, mid-range, or mixed (short/mid)-range parts,
based on its shortest side length $ \tilde{R} $.
Define two distances $ \tilde{R}_{\rm L} $ and $ \tilde{R}_{\rm U} $,
where $ \tilde{R}_{\rm L} < \tilde{R}_{\rm U} $.
The part of the \textit{ab initio} part that a sample falls into is described by}
\begin{equation}
\red{
    \text{part of the \textit{ab initio} PES} =
    \begin{cases}
        \text{short} & \text{if } \tilde{R} \leq \tilde{R}_{\rm L}, \\
        \text{mixed (short/mid)} & \text{if } \tilde{R}_{\rm L} < \tilde{R} < \tilde{R}_{\rm U}, \\
        \text{mid} & \text{if } \tilde{R} \geq \tilde{R}_{\rm U}
    \end{cases}
}
\end{equation}
\noindent
\red{In the mixed (short/mid)-range part,
the energy of a sample is a linear combination of the short-range and mid-range parts.
The fraction of the output that comes from the mid-range part is given by}
\begin{equation}
\red{
    \omega(\tilde{R}; \tilde{R}_{\rm L}, \tilde{R}_{\rm U}) = 
    \begin{cases}
        0                             & \text{if } \tilde{R} \leq \tilde{R}_{\rm L}, \\
        \frac{1}{2} [1 - \cos(\pi k)] & \text{if } \tilde{R}_{\rm L} < \tilde{R} < \tilde{R}_{\rm U}, \\
        1                             & \text{if } \tilde{R} \geq \tilde{R}_{\rm U}
    \end{cases}
}
\end{equation}
\noindent
\red{where $ k = (\tilde{R} - \tilde{R}_{\rm L}) / (\tilde{R}_{\rm U} - \tilde{R}_{\rm L}) $.
In the current PES,
$ \tilde{R}_{\rm L} = 2.2 \ang $ and $ \tilde{r}_{\rm U} = 2.25 \ang $.}

\red{The energies from the short-range and mid-range parts of the PES are already very close
to one another in the mixed (short/mid)-range region,
and so the transition is fairly clean.}

\subsection*{Limitations}

\red{Certain kinds of samples are not treated well with this mixing of, and transition between,
\textit{ab initio} and dispersive energies.
For example,
consider a geometry where
molecules 1 and 2 are less than $ 2.2 \ang $ apart,
molecules 3 and 4 are far away from them (molecules 3 and 4 may be close to one another as well).
If the average side length is between $ \tilde{r}_{\rm L} $ and $ \tilde{r}_{\rm U} $,
the PES calculates a physically unrealistic mix of a short-range exponential wall and a long-range dispersive decay.
If the average side length is greater than $ \tilde{r}_{\rm U} $,
the PES categorizes this sample to be entirely dispersive.
No samples in the training data exist for such geometries,
and the energy produced by the PES for both cases is physically unrealistic.
However,
based on extrapolations of certain geometries in the \textit{hcp} lattice training data,
the four-body energies of such extreme samples are very small
(typically on the order of $ 10^{-3} \wvn $ to $ 1 \wvn $),
and the \textit{ab initio} total two-body energy is very large
(typically on the order of $ 10^{3} \wvn $ to $ 10^{5} \wvn $).
Moreover,
although physically unrealistic,
the transition is still continuously differentiable.}

\red{The short-range exponential decay fit
is based on the observation that,
for small geometries,
the energy of a four-body geometry
varies exponentially as all its side lengths are scaled by the same factor.
The exponential coefficient for the fit
is determined by \textit{ab initio} data that only goes down to side lengths of $ 2.25 \ang $.
The short-range energy predictions made by this PES
for four-body geometries with side lengths much less than $ 2.2 \ang $
are unlikely to be accurate.
In cases where the Boltzmann weight of such small four-body geometries
is non-negligible,
such as in solids at very high densities,
such a limitation might become significant.}

\section{Contributions to the hcp lattice for each geometry}

We supplement the training data using samples
that are based specifically on the geometries of the \textit{hcp} lattice.
The procedure for generating the samples is as follows.
First, we select a reference molecule in the \textit{hcp} lattice.
Then we consider all four-body geometries involving this reference molecule such that
(1) they have at least one side length equal to the lattice constant, and
(2) they have no side length more than twice the lattice constant.
We find that $ 83 $ different four-body geometries satisfy these conditions.
Information about all $ 83 $ geometries is provided in
Table~\ref{tab:supplementary:geometry_contributions}.

\begin{longtable}{|C{1cm}|C{1cm}|C{1.2cm}C{1.2cm}C{1.2cm}C{1.2cm}C{1.2cm}C{1.2cm}|d{3}|}
    \caption{
        Information about particular four-body geometries in the hcp lattice,
        their combinatorial frequency, and their contributions to the average
        four-body interaction potential energy per particle.
        The six side lengths in each geometry are expressed in the permutation
        with the lowest lexicographic order.
        The geometries are then ordered among themselves using their average
        side length,
        with the lexicographic order between their side lengths used as a
        tie-breaker.
        The first column is an ID number assigned to the geometry, for identification.
        The second column contains $ N_c $, the number of times that geometry appears
        under the aforementioned conditions for which they were selected.
        Columns 3 to 8 are the ratio of the six side lengths to the lattice constant.
        Column 9 is the total contribution of the geometry to the average four-body
        interaction potential energy per particle (units of $ \wvn $),
        when the lattice constant is $ 2.2 \ang $.
    } \\

    \hline
    \renewcommand{\arraystretch}{1.3}
    \mbox{ID} & \mbox{$ N_c $} & \mbox{$ r_{12} $} & \mbox{$ r_{13} $} & \mbox{$ r_{14} $} & \mbox{$ r_{23} $} & \mbox{$ r_{24} $} & \mbox{$ r_{34} $} & \multicolumn{1}{c|}{\mbox{$ V_4^{\rm (tot)} $}} \\
    \hline \hline
    \endfirsthead
    
    \hline
    \endfoot
    
    \multicolumn{8}{c}{{\bfseries Table \thetable:} Continued from previous page...} \\
    \hline
    \renewcommand{\arraystretch}{1.3}
    \mbox{ID} & \mbox{$ N_c $} & \mbox{$ r_{12} $} & \mbox{$ r_{13} $} & \mbox{$ r_{14} $} & \mbox{$ r_{23} $} & \mbox{$ r_{24} $} & \mbox{$ r_{34} $} & \multicolumn{1}{c|}{\mbox{$ V_4^{\rm (tot)} $}} \\
    \hline \hline
    \endhead

    \label{tab:supplementary:geometry_contributions}
    $  0 $ & $   8 $ & $                    1 $ & $                    1 $ & $                    1 $ & $                    1 $ & $                    1 $ & $                    1 $ & $  398.3489 $  \\
    $  1 $ & $  48 $ & $                    1 $ & $                    1 $ & $                    1 $ & $                    1 $ & $                    1 $ & $             \sqrt{2} $ & $ 1767.6561 $  \\
    $  2 $ & $  12 $ & $                    1 $ & $                    1 $ & $                    1 $ & $                    1 $ & $                    1 $ & $  \sqrt{\sfrac{8}{3}} $ & $  510.3478 $  \\
    $  3 $ & $  36 $ & $                    1 $ & $                    1 $ & $                    1 $ & $                    1 $ & $                    1 $ & $             \sqrt{3} $ & $ 1673.2181 $  \\
    $  4 $ & $  12 $ & $                    1 $ & $                    1 $ & $             \sqrt{2} $ & $             \sqrt{2} $ & $                    1 $ & $                    1 $ & $  777.0374 $  \\
    $  5 $ & $ 144 $ & $                    1 $ & $                    1 $ & $                    1 $ & $                    1 $ & $             \sqrt{2} $ & $             \sqrt{3} $ & $  938.2455 $  \\
    $  6 $ & $  48 $ & $                    1 $ & $                    1 $ & $                    1 $ & $                    1 $ & $             \sqrt{2} $ & $ \sqrt{\sfrac{11}{3}} $ & $  333.7082 $  \\
    $  7 $ & $  24 $ & $                    1 $ & $                    1 $ & $                    1 $ & $             \sqrt{2} $ & $             \sqrt{2} $ & $  \sqrt{\sfrac{8}{3}} $ & $   26.8497 $  \\
    $  8 $ & $  72 $ & $                    1 $ & $                    1 $ & $                    1 $ & $                    1 $ & $             \sqrt{3} $ & $             \sqrt{3} $ & $  -72.8998 $  \\
    $  9 $ & $  48 $ & $                    1 $ & $                    1 $ & $                    1 $ & $                    1 $ & $  \sqrt{\sfrac{8}{3}} $ & $ \sqrt{\sfrac{11}{3}} $ & $  -34.8335 $  \\
    $ 10 $ & $  48 $ & $                    1 $ & $                    1 $ & $                    1 $ & $                    1 $ & $             \sqrt{3} $ & $ \sqrt{\sfrac{11}{3}} $ & $ -112.8734 $  \\
    $ 11 $ & $  96 $ & $                    1 $ & $                    1 $ & $                    1 $ & $                    1 $ & $             \sqrt{3} $ & $                    2 $ & $ -263.0247 $  \\
    $ 12 $ & $  24 $ & $                    1 $ & $                    1 $ & $             \sqrt{2} $ & $             \sqrt{2} $ & $                    1 $ & $ \sqrt{\sfrac{11}{3}} $ & $   23.2044 $  \\
    $ 13 $ & $  24 $ & $                    1 $ & $                    1 $ & $                    1 $ & $             \sqrt{2} $ & $             \sqrt{2} $ & $                    2 $ & $   31.7028 $  \\
    $ 14 $ & $  24 $ & $                    1 $ & $                    1 $ & $                    1 $ & $                    1 $ & $ \sqrt{\sfrac{11}{3}} $ & $ \sqrt{\sfrac{11}{3}} $ & $ -107.4177 $  \\
    $ 15 $ & $  48 $ & $                    1 $ & $                    1 $ & $                    1 $ & $             \sqrt{2} $ & $             \sqrt{3} $ & $             \sqrt{3} $ & $    3.9874 $  \\
    $ 16 $ & $ 144 $ & $                    1 $ & $                    1 $ & $             \sqrt{2} $ & $             \sqrt{3} $ & $                    1 $ & $             \sqrt{3} $ & $  589.1818 $  \\
    $ 17 $ & $  48 $ & $                    1 $ & $                    1 $ & $             \sqrt{2} $ & $  \sqrt{\sfrac{8}{3}} $ & $                    1 $ & $ \sqrt{\sfrac{11}{3}} $ & $   47.4894 $  \\
    $ 18 $ & $  48 $ & $                    1 $ & $                    1 $ & $                    1 $ & $             \sqrt{2} $ & $             \sqrt{3} $ & $ \sqrt{\sfrac{11}{3}} $ & $   -0.1676 $  \\
    $ 19 $ & $  48 $ & $                    1 $ & $                    1 $ & $             \sqrt{2} $ & $                    1 $ & $             \sqrt{3} $ & $ \sqrt{\sfrac{11}{3}} $ & $   17.5318 $  \\
    $ 20 $ & $  24 $ & $                    1 $ & $                    1 $ & $                    1 $ & $  \sqrt{\sfrac{8}{3}} $ & $             \sqrt{3} $ & $             \sqrt{3} $ & $    8.1791 $  \\
    $ 21 $ & $  12 $ & $                    1 $ & $                    1 $ & $             \sqrt{3} $ & $  \sqrt{\sfrac{8}{3}} $ & $             \sqrt{2} $ & $             \sqrt{2} $ & $   49.4868 $  \\
    $ 22 $ & $   8 $ & $                    1 $ & $                    1 $ & $                    1 $ & $             \sqrt{3} $ & $             \sqrt{3} $ & $             \sqrt{3} $ & $    3.3276 $  \\
    $ 23 $ & $  24 $ & $                    1 $ & $                    1 $ & $             \sqrt{3} $ & $                    1 $ & $             \sqrt{3} $ & $             \sqrt{3} $ & $    2.5747 $  \\
    $ 24 $ & $  24 $ & $                    1 $ & $                    1 $ & $             \sqrt{2} $ & $                    1 $ & $ \sqrt{\sfrac{11}{3}} $ & $ \sqrt{\sfrac{11}{3}} $ & $    1.9198 $  \\
    $ 25 $ & $  48 $ & $                    1 $ & $                    1 $ & $             \sqrt{2} $ & $ \sqrt{\sfrac{11}{3}} $ & $                    1 $ & $ \sqrt{\sfrac{11}{3}} $ & $   78.6803 $  \\
    $ 26 $ & $  48 $ & $                    1 $ & $                    1 $ & $  \sqrt{\sfrac{8}{3}} $ & $             \sqrt{3} $ & $                    1 $ & $ \sqrt{\sfrac{11}{3}} $ & $  147.4124 $  \\
    $ 27 $ & $  48 $ & $                    1 $ & $                    1 $ & $             \sqrt{2} $ & $             \sqrt{2} $ & $             \sqrt{3} $ & $             \sqrt{3} $ & $    5.5033 $  \\
    $ 28 $ & $  12 $ & $                    1 $ & $             \sqrt{2} $ & $             \sqrt{3} $ & $             \sqrt{3} $ & $             \sqrt{2} $ & $                    1 $ & $   56.9016 $  \\
    $ 29 $ & $  48 $ & $                    1 $ & $                    1 $ & $ \sqrt{\sfrac{11}{3}} $ & $             \sqrt{2} $ & $             \sqrt{2} $ & $  \sqrt{\sfrac{8}{3}} $ & $   74.2656 $  \\
    $ 30 $ & $  48 $ & $                    1 $ & $                    1 $ & $  \sqrt{\sfrac{8}{3}} $ & $                    1 $ & $ \sqrt{\sfrac{11}{3}} $ & $ \sqrt{\sfrac{11}{3}} $ & $    0.8395 $  \\
    $ 31 $ & $  48 $ & $                    1 $ & $                    1 $ & $             \sqrt{3} $ & $                    1 $ & $             \sqrt{3} $ & $                    2 $ & $    1.9873 $  \\
    $ 32 $ & $  48 $ & $                    1 $ & $                    1 $ & $             \sqrt{3} $ & $             \sqrt{3} $ & $                    1 $ & $                    2 $ & $  130.0325 $  \\
    $ 33 $ & $  96 $ & $                    1 $ & $                    1 $ & $             \sqrt{3} $ & $             \sqrt{2} $ & $             \sqrt{2} $ & $                    2 $ & $   19.3653 $  \\
    $ 34 $ & $  96 $ & $                    1 $ & $                    1 $ & $             \sqrt{3} $ & $             \sqrt{3} $ & $             \sqrt{2} $ & $             \sqrt{3} $ & $  115.7225 $  \\
    $ 35 $ & $  48 $ & $                    1 $ & $                    1 $ & $             \sqrt{2} $ & $             \sqrt{3} $ & $             \sqrt{3} $ & $ \sqrt{\sfrac{11}{3}} $ & $    2.2902 $  \\
    $ 36 $ & $  48 $ & $                    1 $ & $                    1 $ & $             \sqrt{3} $ & $             \sqrt{2} $ & $             \sqrt{3} $ & $ \sqrt{\sfrac{11}{3}} $ & $    2.2277 $  \\
    $ 37 $ & $  48 $ & $                    1 $ & $             \sqrt{2} $ & $             \sqrt{3} $ & $ \sqrt{\sfrac{11}{3}} $ & $             \sqrt{3} $ & $                    1 $ & $   77.3423 $  \\
    $ 38 $ & $  24 $ & $                    1 $ & $                    1 $ & $             \sqrt{3} $ & $  \sqrt{\sfrac{8}{3}} $ & $             \sqrt{3} $ & $             \sqrt{3} $ & $    6.0805 $  \\
    $ 39 $ & $  24 $ & $                    1 $ & $                    1 $ & $             \sqrt{2} $ & $                    2 $ & $             \sqrt{3} $ & $             \sqrt{3} $ & $    5.6200 $  \\
    $ 40 $ & $  48 $ & $                    1 $ & $                    1 $ & $                    2 $ & $             \sqrt{2} $ & $             \sqrt{3} $ & $             \sqrt{3} $ & $   11.4372 $  \\
    $ 41 $ & $  48 $ & $                    1 $ & $             \sqrt{2} $ & $             \sqrt{3} $ & $             \sqrt{3} $ & $                    2 $ & $                    1 $ & $   19.5027 $  \\
    $ 42 $ & $  24 $ & $                    1 $ & $             \sqrt{2} $ & $             \sqrt{2} $ & $             \sqrt{3} $ & $             \sqrt{3} $ & $  \sqrt{\sfrac{8}{3}} $ & $    1.7976 $  \\
    $ 43 $ & $  48 $ & $                    1 $ & $                    1 $ & $             \sqrt{2} $ & $ \sqrt{\sfrac{11}{3}} $ & $             \sqrt{3} $ & $ \sqrt{\sfrac{11}{3}} $ & $    4.0230 $  \\
    $ 44 $ & $  48 $ & $                    1 $ & $                    1 $ & $ \sqrt{\sfrac{11}{3}} $ & $             \sqrt{2} $ & $             \sqrt{3} $ & $ \sqrt{\sfrac{11}{3}} $ & $    3.2812 $  \\
    $ 45 $ & $  48 $ & $                    1 $ & $                    1 $ & $ \sqrt{\sfrac{11}{3}} $ & $             \sqrt{3} $ & $             \sqrt{2} $ & $ \sqrt{\sfrac{11}{3}} $ & $   31.6711 $  \\
    $ 46 $ & $  48 $ & $                    1 $ & $             \sqrt{2} $ & $             \sqrt{3} $ & $ \sqrt{\sfrac{11}{3}} $ & $ \sqrt{\sfrac{11}{3}} $ & $                    1 $ & $   34.7712 $  \\
    $ 47 $ & $  48 $ & $                    1 $ & $                    1 $ & $ \sqrt{\sfrac{11}{3}} $ & $             \sqrt{3} $ & $  \sqrt{\sfrac{8}{3}} $ & $             \sqrt{3} $ & $   25.8993 $  \\
    $ 48 $ & $  12 $ & $                    1 $ & $  \sqrt{\sfrac{8}{3}} $ & $ \sqrt{\sfrac{11}{3}} $ & $ \sqrt{\sfrac{11}{3}} $ & $  \sqrt{\sfrac{8}{3}} $ & $                    1 $ & $   15.2368 $  \\
    $ 49 $ & $  24 $ & $                    1 $ & $                    1 $ & $                    2 $ & $  \sqrt{\sfrac{8}{3}} $ & $             \sqrt{3} $ & $             \sqrt{3} $ & $    8.0599 $  \\
    $ 50 $ & $  48 $ & $                    1 $ & $             \sqrt{2} $ & $  \sqrt{\sfrac{8}{3}} $ & $             \sqrt{3} $ & $ \sqrt{\sfrac{11}{3}} $ & $             \sqrt{2} $ & $    5.6212 $  \\
    $ 51 $ & $  24 $ & $                    1 $ & $                    1 $ & $             \sqrt{3} $ & $ \sqrt{\sfrac{11}{3}} $ & $             \sqrt{3} $ & $             \sqrt{3} $ & $    8.5996 $  \\
    $ 52 $ & $  48 $ & $                    1 $ & $                    1 $ & $  \sqrt{\sfrac{8}{3}} $ & $             \sqrt{3} $ & $ \sqrt{\sfrac{11}{3}} $ & $ \sqrt{\sfrac{11}{3}} $ & $    1.9287 $  \\
    $ 53 $ & $  96 $ & $                    1 $ & $                    1 $ & $             \sqrt{3} $ & $             \sqrt{3} $ & $             \sqrt{3} $ & $                    2 $ & $    7.0128 $  \\
    $ 54 $ & $  24 $ & $                    1 $ & $                    1 $ & $                    2 $ & $             \sqrt{3} $ & $             \sqrt{3} $ & $             \sqrt{3} $ & $    8.9254 $  \\
    $ 55 $ & $  48 $ & $                    1 $ & $             \sqrt{3} $ & $             \sqrt{3} $ & $             \sqrt{3} $ & $                    2 $ & $                    1 $ & $   24.0374 $  \\
    $ 56 $ & $  24 $ & $                    1 $ & $             \sqrt{2} $ & $             \sqrt{3} $ & $             \sqrt{3} $ & $             \sqrt{2} $ & $ \sqrt{\sfrac{11}{3}} $ & $    2.1125 $  \\
    $ 57 $ & $  48 $ & $                    1 $ & $             \sqrt{2} $ & $             \sqrt{3} $ & $ \sqrt{\sfrac{11}{3}} $ & $             \sqrt{2} $ & $             \sqrt{3} $ & $   12.3522 $  \\
    $ 58 $ & $  48 $ & $                    1 $ & $             \sqrt{2} $ & $                    2 $ & $             \sqrt{3} $ & $             \sqrt{3} $ & $             \sqrt{2} $ & $   13.9028 $  \\
    $ 59 $ & $  48 $ & $                    1 $ & $                    1 $ & $             \sqrt{3} $ & $ \sqrt{\sfrac{11}{3}} $ & $             \sqrt{3} $ & $ \sqrt{\sfrac{11}{3}} $ & $    7.6466 $  \\
    $ 60 $ & $  48 $ & $                    1 $ & $                    1 $ & $ \sqrt{\sfrac{11}{3}} $ & $             \sqrt{3} $ & $             \sqrt{3} $ & $ \sqrt{\sfrac{11}{3}} $ & $    7.0239 $  \\
    $ 61 $ & $  24 $ & $                    1 $ & $             \sqrt{3} $ & $             \sqrt{3} $ & $ \sqrt{\sfrac{11}{3}} $ & $ \sqrt{\sfrac{11}{3}} $ & $                    1 $ & $    7.5321 $  \\
    $ 62 $ & $  48 $ & $                    1 $ & $             \sqrt{2} $ & $             \sqrt{3} $ & $             \sqrt{3} $ & $             \sqrt{3} $ & $             \sqrt{3} $ & $    2.3053 $  \\
    $ 63 $ & $  24 $ & $                    1 $ & $             \sqrt{3} $ & $             \sqrt{3} $ & $             \sqrt{3} $ & $             \sqrt{3} $ & $             \sqrt{2} $ & $    2.4616 $  \\
    $ 64 $ & $  48 $ & $                    1 $ & $  \sqrt{\sfrac{8}{3}} $ & $ \sqrt{\sfrac{11}{3}} $ & $ \sqrt{\sfrac{11}{3}} $ & $ \sqrt{\sfrac{11}{3}} $ & $                    1 $ & $   20.9388 $  \\
    $ 65 $ & $  24 $ & $                    1 $ & $             \sqrt{3} $ & $ \sqrt{\sfrac{11}{3}} $ & $                    2 $ & $             \sqrt{3} $ & $                    1 $ & $   16.0475 $  \\
    $ 66 $ & $  48 $ & $                    1 $ & $             \sqrt{2} $ & $ \sqrt{\sfrac{11}{3}} $ & $             \sqrt{3} $ & $  \sqrt{\sfrac{8}{3}} $ & $             \sqrt{3} $ & $    4.6222 $  \\
    $ 67 $ & $  24 $ & $                    1 $ & $                    1 $ & $  \sqrt{\sfrac{8}{3}} $ & $                    2 $ & $ \sqrt{\sfrac{11}{3}} $ & $ \sqrt{\sfrac{11}{3}} $ & $    1.9854 $  \\
    $ 68 $ & $  12 $ & $                    1 $ & $             \sqrt{3} $ & $                    2 $ & $                    2 $ & $             \sqrt{3} $ & $                    1 $ & $    8.9282 $  \\
    $ 69 $ & $  48 $ & $                    1 $ & $             \sqrt{2} $ & $                    2 $ & $ \sqrt{\sfrac{11}{3}} $ & $             \sqrt{3} $ & $             \sqrt{2} $ & $   16.0005 $  \\
    $ 70 $ & $  48 $ & $                    1 $ & $                    1 $ & $ \sqrt{\sfrac{11}{3}} $ & $             \sqrt{3} $ & $ \sqrt{\sfrac{11}{3}} $ & $ \sqrt{\sfrac{11}{3}} $ & $    2.8300 $  \\
    $ 71 $ & $  48 $ & $                    1 $ & $             \sqrt{3} $ & $             \sqrt{3} $ & $             \sqrt{3} $ & $ \sqrt{\sfrac{11}{3}} $ & $             \sqrt{2} $ & $    4.0850 $  \\
    $ 72 $ & $  12 $ & $                    1 $ & $             \sqrt{3} $ & $             \sqrt{3} $ & $             \sqrt{3} $ & $             \sqrt{3} $ & $  \sqrt{\sfrac{8}{3}} $ & $    0.3714 $  \\
    $ 73 $ & $  48 $ & $                    1 $ & $                    1 $ & $             \sqrt{3} $ & $ \sqrt{\sfrac{11}{3}} $ & $ \sqrt{\sfrac{11}{3}} $ & $                    2 $ & $    2.6144 $  \\
    $ 74 $ & $  24 $ & $                    1 $ & $             \sqrt{2} $ & $             \sqrt{2} $ & $ \sqrt{\sfrac{11}{3}} $ & $ \sqrt{\sfrac{11}{3}} $ & $                    2 $ & $    0.0388 $  \\
    $ 75 $ & $  48 $ & $                    1 $ & $             \sqrt{2} $ & $             \sqrt{3} $ & $ \sqrt{\sfrac{11}{3}} $ & $             \sqrt{3} $ & $ \sqrt{\sfrac{11}{3}} $ & $    1.0870 $  \\
    $ 76 $ & $  48 $ & $                    1 $ & $             \sqrt{2} $ & $             \sqrt{3} $ & $ \sqrt{\sfrac{11}{3}} $ & $ \sqrt{\sfrac{11}{3}} $ & $             \sqrt{3} $ & $    1.1112 $  \\
    $ 77 $ & $  24 $ & $                    1 $ & $                    1 $ & $             \sqrt{3} $ & $                    2 $ & $                    2 $ & $                    2 $ & $    1.2307 $  \\
    $ 78 $ & $  48 $ & $                    1 $ & $  \sqrt{\sfrac{8}{3}} $ & $             \sqrt{3} $ & $ \sqrt{\sfrac{11}{3}} $ & $             \sqrt{3} $ & $             \sqrt{3} $ & $    1.0489 $  \\
    $ 79 $ & $  48 $ & $                    1 $ & $             \sqrt{3} $ & $ \sqrt{\sfrac{11}{3}} $ & $                    2 $ & $             \sqrt{3} $ & $             \sqrt{3} $ & $    0.8478 $  \\
    $ 80 $ & $  24 $ & $                    1 $ & $ \sqrt{\sfrac{11}{3}} $ & $ \sqrt{\sfrac{11}{3}} $ & $ \sqrt{\sfrac{11}{3}} $ & $ \sqrt{\sfrac{11}{3}} $ & $             \sqrt{3} $ & $    0.0701 $  \\
    $ 81 $ & $  24 $ & $                    1 $ & $             \sqrt{3} $ & $             \sqrt{3} $ & $                    2 $ & $                    2 $ & $                    2 $ & $   -0.0329 $  \\
    $ 82 $ & $  96 $ & $                    1 $ & $             \sqrt{3} $ & $ \sqrt{\sfrac{11}{3}} $ & $                    2 $ & $ \sqrt{\sfrac{11}{3}} $ & $ \sqrt{\sfrac{11}{3}} $ & $    0.0262 $  \\
\end{longtable}